\documentclass[11pt,a4paper]{article}

\usepackage[margin=2.2cm]{geometry}
\usepackage[T1]{fontenc}
\usepackage{lmodern}
\usepackage{microtype}
\usepackage{amsmath,amssymb,amsfonts}
\usepackage{graphicx}
\usepackage{grffile}
\usepackage{booktabs}
\usepackage{multirow}
\usepackage{array}
\usepackage{makecell}
\usepackage{threeparttable}
\usepackage{placeins}
\usepackage{afterpage}
\usepackage{catchfilebetweentags}
\usepackage{xcolor}
\usepackage{tikz}
\usepackage{natbib}
\usepackage{hyperref}
\usepackage{cleveref}

\graphicspath{{./}{img/}{outputs/}}

\newcounter{panel}[figure]
\renewcommand{\thepanel}{\alph{panel}}
\newenvironment{subfigure}[2][b]{%
  \refstepcounter{panel}%
  \begin{minipage}[#1]{#2}%
  \centering
  \renewcommand{\caption}[1]{%
    \par\smallskip{\footnotesize(\thepanel)\nobreakspace ##1\par}}%
}{%
  \end{minipage}%
}

\definecolor{Blue}{RGB}{0,0,140}
\hypersetup{colorlinks=true, linkcolor=Blue, citecolor=Blue,
  urlcolor=Blue, filecolor=Blue}

\providecommand{\spacingset}[1]{}

\extrafloats{100}

\title{Tail-Aware Density Forecasting of Locally Explosive Time Series:\\[2pt]
	\large A Neural Network Approach\thanks{We are grateful to Serge Darolles, Christian Francq, Laurent Ferrara, Ga\"elle le Fol, Daniel Velasquez Gaviria, Christian Gouri\'eroux, Alain Hecq, Lo\"ic Henry, Christophe Hurlin, Joann Jasiak, S\'ebastien Laurent, Yannick Le Pen, Gabriele Mingoli, Aryan Manafi Neyazi, Fabrice Rossi and Jean-Michel Zakoian. We also thank the seminar participants at Maastricht University School of Business and Economics, as well as the participants at the 19th CFE conference, the 24th Conference D\'eveloppements R\'ecents de l'Econom\'etrie Appliqu\'ee \`a la Finance (University of Paris Nanterre), the First Workshop in Noncausal Econometrics, the 16th Workshop in Time Series Econometrics (Universidad de Zaragoza), the workshop on Time Series AI in Macroeconomics \& Finance (University of Bristol), the 8th Quantitative Finance and Financial Econometrics conference (Aix-Marseille School of Economics), the 2026 Annual Conference of the International Association for Applied Econometrics (Nova School of Business and Economics) and the 9th International Conference on Econometrics and Statistics (EcoSta 2026, Ryukoku University, Kyoto) for helpful comments and discussions.}}
\author{Elena Dumitrescu\thanks{CRED, University Paris-Panth\'eon-Assas.}
	\and Julien Peignon\thanks{CEREMADE, CNRS UMR 7534, Universit\'e Paris-Dauphine,
		PSL University. Corresponding author
		(\href{mailto:julien.peignon@dauphine.psl.eu}{julien.peignon@dauphine.psl.eu}).}
	\and Arthur Thomas\thanks{LEDa, CNRS, IRD, Universit\'e Paris-Dauphine, Universit\'e PSL.}}
\date{}

\begin{document}

\maketitle

\begin{abstract}
 Mixed causal--noncausal (anticipative) models capture locally explosive dynamics and provide economically meaningful probabilities of continuation and collapse, but forecasting with them remains computationally difficult. We develop a two-stage framework that estimates such an ARMA model and then learns its predictive density using a Mixture Density Network with skewed-$t$ components, tail-aware training weights, and post-hoc calibration. The approach accommodates the heavy tails, asymmetry, and multimodality characteristic of non-causal forecasts while remaining computationally tractable. Monte Carlo experiments and a real-time natural-gas application show substantial improvements over existing density-forecasting methods.
\end{abstract}

\medskip
\noindent\textbf{Keywords:} Forecasting, Noncausal Models, Mixture Density Networks.

\noindent\textbf{JEL classification:} C22, C45, C58.

\clearpage

Stationary causal autoregressive moving average (ARMA) models are inherently mean-reverting: their forecasts converge toward the unconditional mean. Therefore, they have limited ability to represent locally explosive episodes followed by abrupt reversals. Such dynamics are common in asset prices, commodity markets, exchange rates, and other economic variables \citep{HiranoBubbleEconomics2024}. Noncausal autoregressive processes provide an alternative framework. Although these are linear models, their dependence on future innovations generates nonlinear, state-dependent predictive distributions that may be asymmetric, heavy-tailed, and multimodal \citep{GourierouxLocalExplosion2017,FriesConditionalMoments2022,DetruchisForecastingExtreme2025,GourierouxCausalNoncausal2025}. During explosive episodes, these distributions characterize the probabilities of continuation and collapse, making noncausal models particularly relevant for density forecasting and tail-risk assessment.

Noncausal processes also possess a strong theoretical foundation, arising as stationary solutions of rational-expectation models under infinite variance \citep{GourierouxStationaryBubble2020}.
 Empirically, they have been applied to asset prices \citep{FriesMixedCausal2019,GourierouxLocalExplosion2017,GourierouxMisspecificationNoncausal2018,HecqNonCausal2025}, macroeconomic variables \citep{HecqMixedCausal2020,LanneNoncausalAutoregressions2011,DavisNoncausalVector2020}, commodity markets \citep{BlasquesNovelTest2025}, climate extremes \citep{DetruchisForecastingExtreme2025}, green stock prices \citep{GiancateriniBubbleDetection2025}, and exchange rates \citep{CavaliereBootstrappingNoncausal2020}. This literature shows that noncausal specifications can capture nonlinear and locally explosive dynamics that conventional causal models may fail to represent.

Despite their empirical appeal, noncausal models remain difficult to use for forecasting because their predictive densities generally admit no closed-form expression. This difficulty is particularly consequential in extreme states, where the predictive density can become bimodal, reflecting two economically distinct future paths: the bubble either continuing to expand or collapsing. In that case, a point forecast falls between the two modes and describes neither realization.
Existing solutions include simulation-based procedures \citep{LanneOptimalForecasting2012}, Bayesian approaches \citep{NybergForecastingNoncausal2014}, and closed-form functional or semi-parametric estimators of predictive densities \citep{GourierouxFilteringPrediction2016,GourierouxNonlinearFore2026}. The latter substantially reduce the need for computationally intensive simulation and provide point forecasts and prediction intervals from the estimated predictive density. Although these methods have expanded the forecasting toolkit, practical difficulties remain for complex specifications, multiple horizons, and sparsely observed regions of the distribution.

Generic neural density estimators do not directly resolve these challenges. They typically rely on light-tailed components and optimize criteria dominated by frequently observed states. Consequently, rare extreme observations may receive insufficient attention during training \citep{SteiningerDensityWeighting2021,AvelinoResamplingStrategies2024}.
The challenge is therefore to develop a flexible density approximation that preserves the distinctive distributional features of noncausal forecasts while remaining applicable across model specifications and forecast horizons.

We address this challenge with a two-stage density-forecasting framework for mixed causal--noncausal ARMA (MARMA) processes. We first estimate the underlying process and simulate trajectories from the fitted model. We then train a neural density estimator on these trajectories. The network therefore learns the predictive distribution implied by the fitted model rather than directly from the observed series. Because the second stage requires only the ability to simulate from the fitted process, the same density-learning procedure can be applied across specifications, including the general MARMA setting.

The proposed estimator combines three elements. First, the Mixture Density Network (MDN) uses skewed-$t$ rather than Gaussian mixture components to accommodate asymmetry and polynomial tail decay. Second, an adaptive weighting scheme increases the influence of observations associated with extreme conditioning states. Third, a post-hoc recalibration procedure corrects the probabilistic distortion introduced by reweighting.

The paper makes two contributions. First, it provides a unified and computationally tractable method for approximating the predictive densities of MARMA processes. Second, it contributes to the literature on machine learning in economics and finance \citep{AtheyMachineLearning2019,ClarkTailForecasting2023,SalinasDeepAR2020} by adapting neural density estimation to predictive distributions characterized by rare extreme states, asymmetry, and heavy tails.

A distinctive feature of our Monte Carlo analysis is the availability of external theoretical benchmarks. In addition to evaluating forecasts against realized outcomes using proper scoring rules, we examine whether the estimated distributions recover known conditional moments and, in the Cauchy case, the full conditional density. Across these exercises, the MDN ranks among the best-performing methods and most often provides the closest approximation to the theoretical benchmarks. Its advantage is particularly pronounced in the tails.

We then apply the framework to real-time forecasting of U.S. natural gas prices. Holding the estimated noncausal specification fixed, the MDN improves density forecasts relative to the competing estimators. Comparisons with causal Gaussian benchmarks further indicate that the gains reflect the underlying noncausal, heavy-tailed specification rather than the density estimator alone. Although designed for density forecasting, the approach also delivers competitive point forecasts.

More broadly, the paper illustrates how machine-learning methods can complement structurally motivated economic models. The neural network does not replace the noncausal process; it provides a flexible approximation to the predictive distribution implied by the fitted model. This division of labor preserves the model's economic interpretation while making its forecasts operational in practice.\footnote{Replication code is available at \url{https://github.com/JulienPeignon/MARMA-density-forecasting}.}

The remainder of the paper is structured as follows. Section \ref{s3} develops the proposed density-learning framework. Section \ref{s2sec} introduces mixed causal-noncausal ARMA processes and discusses their estimation. Section \ref{s4} presents the Monte Carlo evidence. Section \ref{s5} reports the empirical application to U.S. natural gas prices, and Section \ref{s6} concludes.

\section{A Tail-Aware Mixture Density Network}
\label{s3}

This section presents the Mixture Density Network (MDN) used to approximate the predictive densities implied by causal--noncausal MARMA processes. 
Capturing the characteristic shape of noncausal predictive distributions efficiently requires more than a standard Gaussian-mixture architecture. We therefore combine three ingredients: a flexible mixture representation adapted to heavy-tailed predictive distributions, an adaptive weighting mechanism that increases attention to rare extreme observations during training, and a post-hoc recalibration step that restores probabilistic calibration after reweighting.

\subsection{Mixture Density Network with Skewed t-Distribution Components}

Traditional Mixture Density Networks \citep{BishopMixtureDensity1994} model conditional distributions through mixtures whose parameters are generated by a neural network. In most applications the mixture components are Gaussian.

For noncausal processes, however, the target predictive densities are heavy-tailed. Although Gaussian mixtures are universal approximators, this property is asymptotic and in practice it may require an arbitrarily large number of components. Any finite Gaussian mixture ultimately inherits Gaussian tail decay from its widest component, whereas the predictive density of an $\alpha$-stable MARMA process decays polynomially. Approximation errors therefore become largest precisely in the tail regions that are most relevant for noncausal forecasting.

To better match the shape of noncausal predictive densities, we replace the usual Gaussian components with skewed-$t$ distributions. A skewed-$t$ component accommodates asymmetry and heavy tails directly through its shape and degrees-of-freedom parameters, allowing the mixture to represent tail behaviour parsimoniously. More precisely, our MDN models the conditional density as:
\begin{equation*}
p_h(X_{t+h}|\mathbf{X}_t) = \sum_{j=1}^{K} \pi_{j,h}(\mathbf{X}_t) \cdot f(X_{t+h}; \mu_{j,h}(\mathbf{X}_t), \sigma_{j,h}(\mathbf{X}_t), \xi_{j,h}(\mathbf{X}_t), \nu_{j,h}(\mathbf{X}_t)),
\end{equation*}
with 
$f(y; \mu, \sigma, \xi, \nu) = \frac{2}{\sigma} t\left(\frac{y - \mu}{\sigma}; \nu\right) T\left(\xi \frac{y - \mu}{\sigma} \sqrt{\frac{\nu + 1}{\nu + \left(\frac{y-\mu}{\sigma}\right)^{2}}}; \nu + 1\right)
$ 
the skewed t-distribution probability density function of \cite{AzzaliniDistributionsGenerated2003}, where $t(\cdot; \nu)$ denotes the Student-t density with $\nu$ degrees of freedom, which controls tail thickness, and $T(\cdot; \nu+1)$ the Student-t cumulative distribution function with $\nu+1$ degrees of freedom. This choice also connects our specification to the macro-financial literature, where the skew-$t$ has become the workhorse density for conditional tail risk since \citet{AdrianVulnerableGrowth2019}.
We denote the forecasting horizon by $h$ and let $\mathbf{X}_t = (X_t, X_{t-1}, \ldots, X_{t-L+1})$ be a vector of $L$ consecutive observations up to time $t$. We handle each horizon directly rather than by iteration.

We train a separate network on the pairs $(\mathbf{X}_t, X_{t+h})$, so the network learns the predictive density at horizon $h$ in one step and never chains it from shorter-horizon forecasts. The other parameters are the location $\mu \in \mathbb{R}$, the scale $\sigma > 0$, and the shape $\xi \in \mathbb{R}$, which governs asymmetry.
These predictive densities provide a complete characterization of forecast uncertainty: they can be used to construct prediction intervals, assess tail risk probabilities during explosive episodes, and derive point forecasts. The next subsections describe the network architecture, weighting scheme and recalibration procedure used to estimate these predictive densities. 

\subsection{Network Architecture}

Our network architecture is lightweight and parsimonious, consisting of a fully connected multilayer perceptron (MLP) with $D$ hidden layers of common width $H$ and five parallel output heads, one for each parameter of the skewed t-distribution mixture ($\boldsymbol{\pi}, \boldsymbol{\mu}, \boldsymbol{\sigma}, \boldsymbol{\xi}, \boldsymbol{\nu}$).\footnote{While recurrent architectures such as RNNs or LSTMs could capture additional temporal dependencies, we opt for this simpler feedforward structure to maintain computational efficiency and ease of implementation in applied settings.} 

\begin{figure}[!htbp]
    \centering
    \makebox[\linewidth][c]{%
    \begin{tikzpicture}[x=1.2cm,y=0.7cm,>=stealth,scale=0.76,transform shape]
        \tikzset{
            neuron/.style={circle, draw, minimum size=18pt, inner sep=0pt},
            input neuron/.style={neuron, fill=blue!15},
            hidden neuron/.style={neuron, fill=orange!20},
            output neuron/.style={neuron, fill=green!20},
            missing/.style={draw=none, scale=1.8, text height=0.333cm},
            omitted/.style={->, gray!55}
        }

        \node[input neuron] (I-1) at (0,  2.5) {};
        \node[input neuron] (I-2) at (0,  1.25) {};
        \node[missing]      (I-m) at (0, -0.25) {$\vdots$};
        \node[input neuron] (I-3) at (0, -2.0) {};

        \node[hidden neuron] (H1-1) at (2,  3.0) {};
        \node[hidden neuron] (H1-2) at (2,  1.75) {};
        \node[missing]       (H1-m) at (2,  0.0) {$\vdots$};
        \node[hidden neuron] (H1-3) at (2, -1.75) {};
        \node[hidden neuron] (H1-4) at (2, -3.0) {};

        \node[missing] (M-1) at (4,  3.0) {$\cdots$};
        \node[missing] (M-2) at (4,  1.75) {$\cdots$};
        \node[missing] (M-3) at (4, -1.75) {$\cdots$};
        \node[missing] (M-4) at (4, -3.0) {$\cdots$};

        \node[hidden neuron] (HD-1) at (6,  3.0) {};
        \node[hidden neuron] (HD-2) at (6,  1.75) {};
        \node[missing]       (HD-m) at (6,  0.0) {$\vdots$};
        \node[hidden neuron] (HD-3) at (6, -1.75) {};
        \node[hidden neuron] (HD-4) at (6, -3.0) {};

        \foreach \m [count=\y] in {1,2,3,4,5}
        \node[output neuron] (O-\m) at (8,3.5-1.25*\y) {};

        \draw[<-] (I-1) -- ++(-1.1,0) node[midway,above] {$X_t$};
        \draw[<-] (I-2) -- ++(-1.1,0) node[midway,above] {$X_{t-1}$};
        \draw[<-] (I-3) -- ++(-1.1,0) node[midway,above] {$X_{t-L+1}$};

        \foreach \lab [count=\i] in {
                {$\boldsymbol{\pi}(\mathbf{X}_t) \in [0,1]^K$},
                {$\boldsymbol{\mu}(\mathbf{X}_t) \in \mathbb{R}^K$},
                {$\boldsymbol{\sigma}(\mathbf{X}_t) \in (0,\infty)^K$},
                {$\boldsymbol{\xi}(\mathbf{X}_t) \in (-\xi_{\max},\xi_{\max})^K$},
                {$\boldsymbol{\nu}(\mathbf{X}_t) \in (\nu_{\min},\nu_{\max})^K$}
            }
        \draw[->] (O-\i) -- ++(1.0,0) node[right] {\lab};

        \foreach \i in {1,2,3}
        \foreach \j in {1,2,3,4}
        \draw[->] (I-\i) -- (H1-\j);

        \foreach \i in {1,2,3,4}
        \foreach \j in {1,2,3,4}
        \draw[omitted] (H1-\i) -- (M-\j);

        \foreach \i in {1,2,3,4}
        \foreach \j in {1,2,3,4}
        \draw[omitted] (M-\i) -- (HD-\j);

        \foreach \i in {1,2,3,4}
        \foreach \j in {1,...,5}
        \draw[->] (HD-\i) -- (O-\j);

        \node[align=center, above] at (0, 3.3) {Input\\layer};
        \node[align=center, above] at (2, 3.4) {Hidden\\layer 1};
        \node[align=center, above] at (4, 3.4) {Hidden layers\\$2,\ldots,D-1$};
        \node[align=center, above] at (6, 3.4) {Hidden\\layer $D$};
        \node[align=center, above] at (8, 3.3) {Output\\heads};

        \node[align=center, below] at (0, -3.0) {$L$ lags};
        \node[align=center, below] at (2, -4.0) {$H$ units};
        \node[align=center, below] at (6, -4.0) {$H$ units};
        \node[align=center, below] at (8, -3.2) {$K$ values each};
    \end{tikzpicture}}
    \caption{Architecture of the Mixture Density Network}
    \label{fig:mdn_architecture}
    \parbox{\textwidth}{\scriptsize
    \textit{Notes:} The network maps the $L$ conditioning lags $\mathbf{X}_t = (X_t, \ldots, X_{t-L+1})$ to the parameters of a $K$-component skewed $t$ mixture, through $D$ fully connected hidden layers of common width $H$ with ReLU activations. Each of the five output heads returns one parameter vector of length $K$, with the activation reported in Section~\ref{s3} enforcing its support. The depth $D$, the width $H$, the number of components $K$ and the number of lags $L$ are selected by Bayesian optimization.}
\end{figure}

The hidden layers use rectified linear unit (ReLU) activations, a standard choice in deep learning due to its computational efficiency and ability to mitigate vanishing gradient issues \citep{GoodfellowDeepLearning2016}:
\begin{align*}
\mathbf{z}^{(0)}(\mathbf{X}_t) &= \mathbf{X}_t \in \mathbb{R}^L, \\
\mathbf{z}^{(d)}(\mathbf{X}_t) &= \text{ReLU}\left(\mathbf{W}^{(d-1)}\mathbf{z}^{(d-1)}(\mathbf{X}_t) + \mathbf{b}^{(d-1)}\right) \in \mathbb{R}^{H}, \quad d = 1, \ldots, D,
\end{align*}
where $\mathbf{W}^{(0)} \in \mathbb{R}^{H \times L}$, $\mathbf{W}^{(d)} \in \mathbb{R}^{H \times H}$ for $d \geq 1$, $\mathbf{b}^{(d)} \in \mathbb{R}^{H}$, and $\text{ReLU}(x) = \max(0, x)$ is applied element-wise. Dropout is optionally applied after each hidden activation. The five output heads map the final hidden representation $\mathbf{z}^{(D)}(\mathbf{X}_t)$ to the mixture parameters:
\begin{align*}
\boldsymbol{\pi}(\mathbf{X}_t) &= \text{Softmax}(\mathbf{W}_\pi \mathbf{z}^{(D)}(\mathbf{X}_t) + \mathbf{b}_\pi) \in [0,1]^K, \; \sum_{j=1}^K \pi_j = 1 \\
\boldsymbol{\mu}(\mathbf{X}_t) &= \mathbf{W}_\mu \mathbf{z}^{(D)}(\mathbf{X}_t) + \mathbf{b}_\mu \in \mathbb{R}^K\\
\boldsymbol{\sigma}(\mathbf{X}_t) &= \text{Softplus}(\mathbf{W}_\sigma \mathbf{z}^{(D)}(\mathbf{X}_t) + \mathbf{b}_\sigma) \in (0, \infty)^K\\
\boldsymbol{\xi}(\mathbf{X}_t) &= \xi_{\max} \times \tanh(\mathbf{W}_\xi \mathbf{z}^{(D)}(\mathbf{X}_t) + \mathbf{b}_\xi) \in (-\xi_{\max}, \xi_{\max})^K\\
\boldsymbol{\nu}(\mathbf{X}_t) &= \nu_{\min} + (\nu_{\max} - \nu_{\min}) \times \text{Sigmoid}(\mathbf{W}_\nu \mathbf{z}^{(D)}(\mathbf{X}_t) + \mathbf{b}_\nu) \in (\nu_{\min}, \nu_{\max})^K,
\end{align*}
where $K$ is the number of mixture components, and each output head has weight matrix $\mathbf{W}_\bullet \in \mathbb{R}^{K \times H}$ and bias $\mathbf{b}_\bullet \in \mathbb{R}^K$. Figure~\ref{fig:mdn_architecture} sketches the resulting network.
The softmax function, $\text{Softmax}(\mathbf{x})_j = e^{x_j}/\sum_{i=1}^K e^{x_i}$, maps an unconstrained $K$-vector to valid probability weights $(\pi_j)_{j=1}^K$ summing to one. The softplus function, $\text{Softplus}(x) = \log(1 + e^x)$, keeps the scale strictly positive. The logistic function, $\text{Sigmoid}(x) = 1/(1+e^{-x})$, maps the degrees of freedom onto $(\nu_{\min}, \nu_{\max}) = (2, 100)$, and the hyperbolic tangent maps the skewness onto $(-\xi_{\max}, \xi_{\max})$ with $\xi_{\max} = 6$. The lower bound $\nu_{\min} = 2$ ensures that the conditional predictive densities have a finite variance.\footnote{The heaviest-tailed noncausal process we study has Cauchy errors, for which the unconditional mean is undefined. Nonetheless, its conditional variance exists \citep{FriesConditionalMoments2022}, so the conditional predictive density is square integrable.} The remaining bounds, $\nu_{\max}$ and $\xi_{\max}$, stabilize training while leaving enough range for expressivity.\footnote{$\nu_{\max} = 100$ is close to Gaussian tails. The Student-$t$ excess kurtosis falls to about $0.06$ at $\nu = 100$ versus $0$ for the Gaussian. $\xi_{\max} = 6$ lets the component be highly asymmetric without becoming degenerate. For this parameterization, it places about 95\% of the density mass on one side of the mode.}

The model is implemented in PyTorch with weights initialized via the Kaiming uniform scheme \citep{HeDelvingDeep2015}. The $\boldsymbol{\nu}$ and $\boldsymbol{\xi}$ heads are the exception. We initialize them at $\nu = 5$ and $\xi = 0$, so that every component starts symmetric and moderately heavy-tailed. Subsequent departures from these initial values are data-driven. We train the network by mini-batch stochastic gradient descent, minimizing the negative log-likelihood of the mixture density.\footnote{A mini-batch $\mathcal{B} \subset \mathcal{D}$ is a small random subset of training observations used to compute gradient updates at each iteration of the stochastic gradient descent.} We use the rectified Adam optimizer \citep{LiuVarianceAdaptive2020}, a variant of Adam \citep{KingmaAdamMethod2015} that removes the excessive variance of the adaptive learning rate in the early iterations, together with decoupled weight decay \citep{LoshchilovDecoupledWeight2019}.

We do not fix the architecture \emph{a priori}. Bayesian optimization \citep{AkibaOptunaNext2019} selects the depth $D$, the width $H$, the number of mixture components $K$, the dropout rate and the learning rate, and we retain the configuration minimizing the negative log-likelihood on a held-out validation sample. Section~\ref{s4} details the procedure and the search spaces (Table~\ref{tab:hyperparams}).

\subsection{Adaptive Weighting Function}

Real asset prices naturally exhibit imbalance between normal market conditions and extreme events, yet learning algorithms prioritize frequent observations over rare tail events that are crucial for risk assessment. To address this, we combine resampling strategies \citep{AvelinoResamplingStrategies2024} with cost-sensitive learning \citep{SteiningerDensityWeighting2021}, which encourage the MDN to focus on extreme events during training.

Let $\mathcal{D}$ denote the training sample of size $T = |\mathcal{D}|$, and let the subset of tail events in the training sample be given by
$\mathcal{E} = \{t \in \mathcal{D} : X_{t} < \ell \text{ or } X_{t} > u\}.$
The weighting scheme therefore emphasizes forecasts originating from extreme states: membership in $\mathcal{E}$ is determined by the current observation $X_t$, rather than by the future outcome $X_{t+h}$. In particular, $\mathcal{E}$ depends only on the scalar $X_t$, not on the entire conditioning vector $\mathbf X_t$. The boundaries $\ell$ and $u$ are determined using the generalized boxplot methodology of \cite{BruffaertsGeneralizedBoxplot2014}.\footnote{
$\ell = \xi^{\text{TGH}}_{\delta/2}, \quad u = \xi^{\text{TGH}}_{1-\delta/2},$ for a fixed detection rate $\delta$ set to $0.05$ throughout the paper,
where $\xi^{\text{TGH}}$ is the empirical Tukey g-h CDF fitted to a rank-preserving transformation of the observed data to accurately capture skewness and tail heaviness of the data.
}
Then, the adaptive weight function is given by:
\begin{equation}
\label{weight}
w_t = \begin{cases}
\sqrt{\frac{T}{|\mathcal{E}|}}, & \text{if } t \in \mathcal{E} \\
1, & \text{otherwise}.
\end{cases}
\end{equation}
The weighting mechanism operates at two levels during training, which allows the model to focus on rare tail events without requiring modifications to the underlying MDN architecture.
First, during mini-batch construction, we give more weight to rare cases by employing a weighted random sampling with replacement approach. Under this scheme, the probability that observation $t$ is sampled in a mini-batch $\mathcal{B}$ is given by
$P(t) = {w_t}/ {\sum_{i=1}^{T} w_i}.$
Second, within each mini-batch $\mathcal{B}$, we apply cost-sensitive learning through a weighted loss function. At each gradient step, the model parameters $\boldsymbol{\theta}$ are updated by minimizing the weighted negative log-likelihood,
\begin{equation*}
\mathcal{L}(\boldsymbol{\theta}) = -\frac{1}{|\mathcal{B}|} \sum_{j \in \mathcal{B}} w_j \log p_h(X_{j+h}|\mathbf{X}_j; \boldsymbol{\theta}),
\end{equation*}
where $j$ indexes observations in mini-batch $\mathcal{B}$ and the weights in \eqref{weight} are rescaled to have unit mean over the training sample, so that the loss remains on the scale of an unweighted average log-likelihood. This dual application of weights, both in sampling and loss computation, results in an effective weight of $w_t^2 = T/|\mathcal{E}|$ for tail events. This justifies the square-root scaling in \eqref{weight} to achieve the desired inverse proportion weighting.

\subsection{Post-hoc Calibration}\label{sect:calibration}

By reweighting the training distribution to emphasize extreme events, we train the model on a distribution that differs from the data-generating process. Specifically, when we assign higher weights to tail observations, the model learns to predict conditional distributions $\hat{p}_h(X_{t+h}|\mathbf{X}_t)$ that reflect this reweighted empirical distribution rather than the original empirical distribution. This shift causes systematic biases: the model overestimates the probability of extreme events across the entire input space, leading to miscalibrated predictions when evaluated with respect to the original, unweighted distribution.

To correct for miscalibration, we use the method of \cite{DeyReCalibrating2021} based on the uniformity property of the Probability Integral Transform (PIT), $\text{PIT}_h(X_{t+h}, \mathbf{X}_t) = \int_{-\infty}^{X_{t+h}} \hat{p}_h(z|\mathbf{X}_t) \, dz = \hat{F}_h(X_{t+h}|\mathbf{X}_t),$ where $\hat{F}_h$ is the predicted cumulative distribution function.
This recalibration procedure is applied after the initial model training on $\mathcal{D}$ and requires a distinct calibration dataset, $\mathcal{D}_{\text{cal}}$. First, we use the trained model to compute the PIT values for all observation pairs $(X^{\text{cal}}_{t+h}, \mathbf{X}^{\text{cal}}_t) \in \mathcal{D}_{\text{cal}}$. Then, the conditional distribution of PIT values is estimated as follows: for each threshold $\tau$ on a grid $\mathcal{G} \subset [0,1]$, a separate XGBoost classifier \citep{ChenXgboostScalable2016} is trained to predict the binary outcome $\mathbf{1}\{\hat{F}_h(X_{t+h}|\mathbf{X}_t) \leq \tau\}$ from the calibration features $\mathbf{X}^{\text{cal}}_t$. The predicted probability from each classifier provides an estimate of
\begin{equation*}
\hat{\beta}_h(\tau|\mathbf{X}^{\text{cal}}_t) = \mathbb{P}(\hat{F}_h(X^{\text{cal}}_{t+h}|\mathbf{X}^{\text{cal}}_t) \leq \tau | \mathbf{X}^{\text{cal}}_t), \quad \tau \in [0,1].
\end{equation*}
Intuitively, $\hat{\beta}_h(\tau|\mathbf{X}^{\text{cal}}_t)$ measures the empirical frequency with which the model's predicted CDF, evaluated at the true outcome, falls below $\tau$ for observations with similar features. Perfect calibration corresponds to $\hat{\beta}_h(\tau|\mathbf{X}^{\text{cal}}_t) = \tau$ for all $\tau$, which would indicate that the PIT values are uniformly distributed conditionally on $\mathbf{X}^{\text{cal}}_t$.

Once the correction function $\hat{\beta}_h$ is learned on $\mathcal{D}_{\text{cal}}$, it can be applied to any new observation pair $(X^{\text{test}}_{t+h}, \mathbf{X}^{\text{test}}_t) \in \mathcal{D}_{\text{test}}$ from the test set or future data. The final recalibrated PDF, $\hat{p}^{\text{recal}}_{h}$, is obtained by applying the correction and then renormalizing,
\begin{equation*}
\hat{p}^{\text{recal}}_{h}(X^{\text{test}}_{t+h}|\mathbf{X}^{\text{test}}_t) = \frac{c_h(X^{\text{test}}_{t+h}|\mathbf{X}^{\text{test}}_t) \cdot \hat{p}_h(X^{\text{test}}_{t+h}|\mathbf{X}^{\text{test}}_t)}{\int c_h(z|\mathbf{X}^{\text{test}}_t) \cdot \hat{p}_h(z|\mathbf{X}^{\text{test}}_t) \, dz},
\label{pdens}
\end{equation*}
with correction factor $c_h(X^{\text{test}}_{t+h}|\mathbf{X}^{\text{test}}_t) = \frac{d\hat{\beta}_h}{d\tau}\bigg|_{\tau = \hat{F}_h(X^{\text{test}}_{t+h}|\mathbf{X}^{\text{test}}_t)}.$
The correction factor is derived from a change-of-variables formula which ensures that the recalibrated density produces uniformly distributed PIT values. Since the classifiers are fitted independently at each threshold, their outputs are not necessarily monotone in $\tau$, whereas $\hat{\beta}_h(\cdot|\mathbf{X}_t)$ is a conditional distribution function and must be non-decreasing. Following \cite{DeyReCalibrating2021}, we therefore fit a monotone I-spline expansion to $\tau \mapsto \hat{\beta}_h(\tau|\mathbf{X}_t)$ over the grid $\mathcal{G}$ and take $c_h$ as its analytical derivative.


\section{Mixed Causal-Noncausal ARMA Processes}
\label{s2sec}

The MDN introduced in Section~\ref{s3} learns the predictive density implied by an estimated causal--noncausal process rather than directly from the observed series. The first stage of the forecasting procedure therefore consists in specifying and estimating a model capable of reproducing locally explosive dynamics. We use the class of mixed causal-noncausal ARMA processes and describe below their properties and statistical inference.

\subsection{Definition and Properties}
\label{s2}
Mixed causal-noncausal ARMA processes are appealing because, despite remaining linear in their innovations, they can generate conditional predictive distributions exhibiting strong asymmetry, heavy tails and, in some regions of the state space, multimodality.
Let $X_t$ $(t = 0, \pm 1, \pm 2, \ldots)$ be a stochastic process generated by
\begin{equation}
\label{eq_recur}
  \psi(F)\phi(B)X_t = \theta(F)\eta(B)\varepsilon_t,
\end{equation}
where $F$ (resp. $B = F^{-1}$) denotes the forward (resp. backward) operator,
$\phi(z)=1 - \phi_{1}z - \ldots - \phi_{p} z^p$, $\psi(z)= 1 - \psi_{1}z - \ldots - \psi_{q} z^q$, $\eta(z)= 1 - \eta_{1}z - \ldots - \eta_{r} z^r$, and $\theta(z)=1 - \theta_{1}z - \ldots - \theta_{s} z^s$ are polynomials with roots outside the unit circle, and $(\varepsilon_t)_{t\in\mathbb{Z}}$ is a sequence of i.i.d. variables.
The polynomials in $B$ determine the causal component of the representation, while the polynomials in $F$ determine its anticipative component.
Equation \eqref{eq_recur} admits a unique stationary solution, called a $MARMA(p,q,r,s)$, if $\phi(z) \neq 0$ and $\psi(z) \neq 0$ for all $\mid z \mid \leq 1$, and if $\psi$ (resp. $\phi$) has no common root with $\theta$ (resp. $\eta$). The four orders thus run causal autoregressive, noncausal autoregressive, causal moving average, noncausal moving average. The solution is said to be noncausal if $q\geq 1$ and purely noncausal for $q\geq1$ and $p=0$. If $\eta(B)=\theta(F)=1$, the process simplifies to a MAR$(p,q)$. The purely noncausal AR(1) of Equation~\eqref{eq:mar01} below is therefore a MAR$(0,1)$.
Identification of the causal and noncausal roots relies on departures from Gaussianity. A sufficient condition for the identification of the model in (\ref{eq_recur}) is that the innovations are i.i.d. and non-Gaussian \citep{RosenblattGaussianNon2000}.

The literature has considered both $\alpha$-stable and $t$-distributed innovations, with the choice of innovation distribution largely determining the forecasting tools that can be employed. We focus on $\alpha$-stable innovations for two reasons, though our approach applies equally to $t$-distributed innovations. First, the family is flexible enough to encompass distributions ranging from the Gaussian ($\alpha=2$) to the Cauchy ($\alpha=1$) for $\beta = 0$. Second, and more importantly for our objective, the $\alpha$-stable framework provides theoretical benchmarks against which predictive densities can be evaluated. Specifically, \cite{GourierouxLocalExplosion2017} derived a closed-form expression for the predictive conditional density of a purely noncausal MAR$(0,1)$ when $\varepsilon_t \overset{\text{i.i.d.}}{\sim}$ Cauchy.
Under the corresponding existence conditions, \citet{FriesConditionalMoments2022} derive closed-form expressions for predictive conditional moments up to order four. In our analysis, we consider the integer orders $m\in\{1,\ldots,4\}$ satisfying $m<2\alpha+1$.
These theoretical results play an important role in our Monte Carlo analysis in Section~\ref{s4}, where they provide external benchmarks against which the MDN forecasts can be evaluated. 

\subsection{Statistical Inference}
\label{sec:param-estimation}

The MDN is trained on trajectories simulated from an estimated anticipative model rather than directly on the observed series. Applying the forecasting framework therefore requires two ingredients: an estimate of the underlying MARMA process and the ability to generate simulated trajectories from it.
We briefly review the available estimation methods before describing the choices adopted in Sections~\ref{s4} and~\ref{s5}.


For the subclass of MAR$(p,q)$ processes, three estimation strategies coexist in the literature. The first is (approximate) maximum likelihood (ML), which adapts the method of \citet{LanneNoncausalAutoregressions2011} to $\alpha$-stable innovations \citep[see also][]{AndrewsMaximumLikelihood2009} and estimates the autoregressive coefficients jointly with the parameters of the innovation distribution. The second is the Generalized Covariance (GCov) estimator \citep{GourierouxGeneralizedCovariance2023}, a semi-parametric method that minimizes a portmanteau statistic based on the autocovariances of nonlinear transformations of the model residuals. It does not require specifying the innovation distribution. One estimates the parameters of that distribution separately, for instance by fitting the characteristic-function-based estimator of \citet{NolanUnivariateStable2020} to the filtered residuals. The third is a frequency-domain minimum-distance estimator based on higher-order spectra \citep{VelascoLobatoFrequency2018}, later extended to the noncausal-noninvertible setting by \citet{HecqNonCausal2025}.

The distinction between these methods is particularly relevant under the $\alpha$-stable innovations considered in this paper. GCov is semi-parametric, but its asymptotic theory relies on nonlinear transformations of the residuals admitting finite fourth-order moments \citep{GourierouxGeneralizedCovariance2023}, whereas an $\alpha$-stable variable possesses finite moments only up to order $\alpha$. Approximate ML faces no such restriction, since estimation proceeds directly from the $\alpha$-stable likelihood. These considerations are especially important because the first-stage estimator directly determines the trajectories on which the MDN is trained.

Accordingly, approximate ML is adopted as our baseline estimation method throughout the paper, while GCov is used primarily for robustness analysis. 
Note that for the general MARMA$(p,q,r,s)$ case, only the frequency-domain spectral approach \citep{VelascoLobatoFrequency2018, HecqNonCausal2025} is currently available in the literature. We complement it with our own approximate ML estimator for the MARMA$(1,1,1,1)$ specification used in Section~\ref{s4}, whose likelihood is available in closed form and can therefore be maximized directly. Extending ML to the general MARMA$(p,q,r,s)$ case, and establishing whether the GCov estimator can accommodate moving-average components, remain open questions that fall outside the scope of this paper.

Whichever estimator is used, selecting the orders $(p,q)$ of an unknown MAR process is itself nontrivial: standard tools such as the sample partial autocorrelation function (PACF) or information criteria are typically justified under finite-variance innovations, whereas $\alpha$-stable innovations have infinite variance for $\alpha < 2$. \citet{DavisResnickLimitTheory1986} derive the limit theory of the sample autocorrelation and partial autocorrelation functions of linear processes under regularly-varying, infinite-variance innovations. \citet{AndrewsMaximumLikelihood2009} note that sample correlations remain informally useful for suggesting a specification even in that setting. \citet{KnightConsistencyAkaike1989} further shows that the AIC remains a consistent order-selection criterion for infinite-variance autoregressions, unlike the BIC, for which no analogous result is available.

Taken together, these results suggest that standard identification tools remain informative even in the infinite-variance setting. We rely on this evidence, together with the portmanteau specification test of \citet{JasiakGcovPortmanteau2026}, when selecting the order of the MAR process fitted to the natural gas series in the empirical application in Section \ref{s5}.

The role of model identification differs across our two illustrations. In the Monte Carlo experiments, the orders $(p,q,r,s)$ are assumed known so that forecast accuracy can be assessed independently of specification uncertainty.
In the empirical application of Section~\ref{s5}, where the true order is unknown, we first select the specification using the GCov estimator together with the portmanteau test of \citet{JasiakGcovPortmanteau2026}, since GCov does not require the innovation distribution to be specified at that stage. For the retained specification, we then report both the GCov and the ML estimates of the MAR coefficients and of the $\alpha$-stable parameters; the two methods deliver similar conclusions. To avoid redundancy, the forecasting results reported thereafter rely on the ML estimates only, for the reasons given above.

The purpose of this first-stage estimation is not inference on the MARMA parameters themselves, but the construction of simulated trajectories from which the MDN learns the predictive density associated with the fitted anticipative process. Nothing in the MDN architecture or training procedure is tied to the $\alpha$-stable assumption. We adopt $\alpha$-stable innovations because they provide both a flexible representation of heavy-tailed behaviour and theoretical benchmarks for forecast evaluation through the results of \cite{GourierouxLocalExplosion2017} and \cite{FriesConditionalMoments2022}. The same learning framework could be applied to any innovation distribution from which trajectories can be simulated.

Throughout, standard errors for the ML estimates are obtained from the inverse of the observed information matrix, computed by finite differences of the gradient and Hessian of the log-likelihood, evaluated in an unconstrained reparameterization of the parameters constrained to a subset of $\mathbb{R}$ (e.g.\ $\alpha$, $\sigma$). This is exact for the $\alpha$-stable innovation parameters, which retain the traditional $\sqrt{n}$ rate of convergence and are asymptotically normal, independently of the autoregressive estimators \citep{AndrewsMaximumLikelihood2009}. For the autoregressive coefficients, however, this normal approximation is only indicative: under $\alpha$-stable innovations their ML estimators are $n^{1/\alpha}$-consistent and converge to the maximizer of a random function whose limiting distribution is non-Gaussian and analytically intractable \citep{AndrewsMaximumLikelihood2009,CavaliereBootstrappingNoncausal2020}. The bootstrap procedures proposed by these authors to approximate this limiting distribution are established only for purely causal or purely noncausal processes; since our retained specification is the mixed MAR$(1,1)$, they do not directly apply here, and we are not aware of an extension covering the mixed case. We therefore continue to report Hessian-based standard errors for the autoregressive coefficients, while noting that, absent a bootstrap procedure valid for mixed causal-noncausal models, they should be interpreted with caution. For the GCov estimator, standard errors for the autoregressive coefficients are instead computed analytically from the formula given in Corollary~1 of \citet{GourierouxGeneralizedCovariance2023}, which, given our choice of instrument $g_k$, requires the first-order derivative of $\gamma(h;\theta_0)$ with respect to $\theta_0$; we again approximate this derivative by finite differences. Because GCov estimates the innovation distribution in a second step, from the filtered residuals, the standard errors of the $\alpha$-stable parameters in that case are instead those returned by the characteristic-function-based estimator of \citet{NolanUnivariateStable2020}.

\section{Monte Carlo Simulations}
\label{s4}

This section evaluates the ability of the MDN to recover the predictive density implied by estimated anticipative processes. We consider several MARMA specifications for which existing theoretical results provide external benchmarks, allowing predictive-density accuracy to be assessed along multiple dimensions.

\subsection{Simulation Design}
\label{SD}

\subsubsection{Data-Generating Processes}

We generate time series of length 2,000 from the following MARMA data generating processes \citep{DetruchisLaurentSeries2026}:
\begin{align}
\text{MAR(0,1):} \quad & (1-0.9F)X_t = \varepsilon_t, \label{eq:mar01} \\
\text{MAR(0,2):} \quad & (1-0.7F-0.1F^2)X_t = \varepsilon_t, \label{eq:mar02} \\
\text{MAR(1,1):} \quad & (1-0.9F)(1-0.1B)X_t = \varepsilon_t, \\
\text{MARMA(1,1,1,1):} \quad & (1-0.9F)(1+0.3B)X_t = (1+0.4F)(1-0.3B)\varepsilon_t,
\end{align}
where $\varepsilon_t \overset{\text{i.i.d.}}{\sim} \mathcal{S}(\alpha, 0, 0.5, 0)$.\footnote{The parameter values are chosen by following \cite{DetruchisForecastingExtreme2025} for the three MAR models and based on \cite{FriesConditionalMoments2022} for the MARMA specification. All specifications satisfy the stationarity conditions.} Together, these specifications span the main classes considered in the anticipative literature, ranging from the benchmark MAR(0,1) process for which analytical density results are available to higher-order MAR and MARMA models exhibiting increasingly complex predictive distributions.
An additional calibration set $\mathcal{D}_{\text{cal}}$ of 1,000 observations, generated by the same data-generating process, is used to perform the local recalibration procedure described in Section~\ref{sect:calibration}.
We consider $\alpha \in \{1.0,1.2,1.4,1.6,1.8\}$, thereby evaluating performance across progressively lighter-tailed innovation distributions.

\subsubsection{Competing Forecasting Methods}

The comparison combines methods specifically developed for anticipative processes with general-purpose conditional density estimators.\footnote{For a thorough presentation of these methods, see the Online Appendix.}


Two competitors are specifically designed for anticipative processes: the simulation-based method of \citet{LanneOptimalForecasting2012} and the kernel-based semi-nonparametric estimator of \citet{GourierouxNonlinearFore2026}.\footnote{In the univariate case retained here, this estimator is strictly equivalent to the predictive density derived by \cite{GourierouxFilteringPrediction2016} for univariate mixed processes, the two differing only in the way the stationary density of the noncausal state is estimated. We therefore report a single benchmark for both. The Online Appendix details the equivalence.}
The remaining two are general-purpose conditional density estimators, not tied to any specific anticipative structure: the Nadaraya-Watson kernel density estimator of \citep{RosenblattConditionalProbability1969}, and the learning-based FlexZBoost method of \cite{IzbickiConvertingHigh2017} and \cite{DalmassoConditionalDensity2020}.

To ensure comparability, the recalibration procedure of Section~\ref{sect:calibration} is applied to every forecasting method, not only to the MDN. All five predictive densities are recalibrated on the same sample $\mathcal D_{\mathrm{cal}}$ using the same local PIT procedure. Differences in forecast performance therefore reflect the underlying density estimators rather than unequal calibration.

\subsubsection{Accounting for Parameter Uncertainty}

An important feature of our design is that the parameters used throughout are not the generating ones, but estimates carrying the sampling uncertainty of a realistically short sample. Each experiment starts from an estimation sample of 300 observations drawn from the true data-generating process, a length comparable to the financial series to which these models are applied in practice. The MAR(MA) coefficients and the $\alpha$-stable parameters are estimated exclusively from this sample. These estimates are then used to generate the 2,000-observation training trajectory and the 1,000-observation calibration sample, and are supplied to the two model-based benchmarks. Only the evaluation sample is generated from the true data-generating process. All methods are therefore exposed to the same first-stage sample information and estimation protocol, while forecast accuracy is evaluated against the process that generated the data.


\subsection{Bimodality Analysis and Sampled Trajectories} 
\label{sec:bimodality}

Before comparing forecast accuracy, we examine whether the competing methods recover one of the central theoretical predictions of anticipative processes: the bimodality of the conditional predictive density.

\subsubsection{Bimodality of the Conditional Predictive Density}
\label{sec:grid}

For the class of MAR processes with a single anticipative root considered by \citet{FriesConditionalMoments2022}, extreme conditioning values induce a bimodal predictive distribution; see also \citet{DetruchisForecastingExtreme2025} for extensions to a broader class of processes. 
This reflects the continuation-versus-collapse structure of anticipative forecasts in extreme states, with probabilities $|\psi_1|^{\alpha h}$ and $1-|\psi_1|^{\alpha h}$ respectively (see Section 2.2 in the Online Appendix for a discussion). The two modes represent continuation and collapse: one lies closer to the center of the marginal distribution, while the other corresponds to continuation away from the center. 

To assess the ability of each method to capture this bimodality, the procedure relies on a grid of 5,000 equispaced conditioning values $\{x_1, \ldots, x_{5000}\}$ spanning the quantile range $[q_{0.01}, q_{0.99}]$, where $q_\tau$ denotes the $\tau$-quantile of the theoretical marginal distribution of the process considered. For each point $x_i$ on the grid, we estimate the conditional predictive density $\hat{p}_h(X_{t+h} \mid X_t = x_i)$ using each of the competing methods and visually assess their ability to recover the bimodal structure.

\begin{figure}[!htbp]
    \centering
    \begin{subfigure}{\linewidth}
        \centering
        \includegraphics[width=0.85\linewidth,height=0.14\textheight,keepaspectratio]{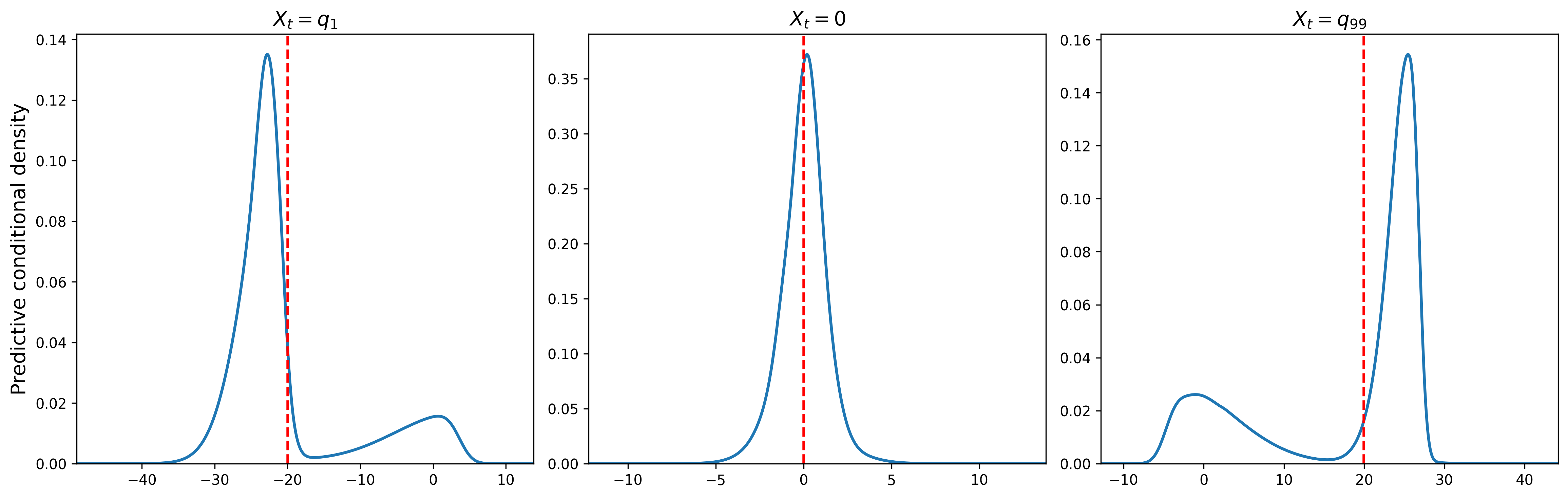}
        \caption{Mixture Density Network}
    \end{subfigure}
\vspace{-0.3em}

    \begin{subfigure}{\linewidth}
        \centering
        \includegraphics[width=0.85\linewidth,height=0.14\textheight,keepaspectratio]{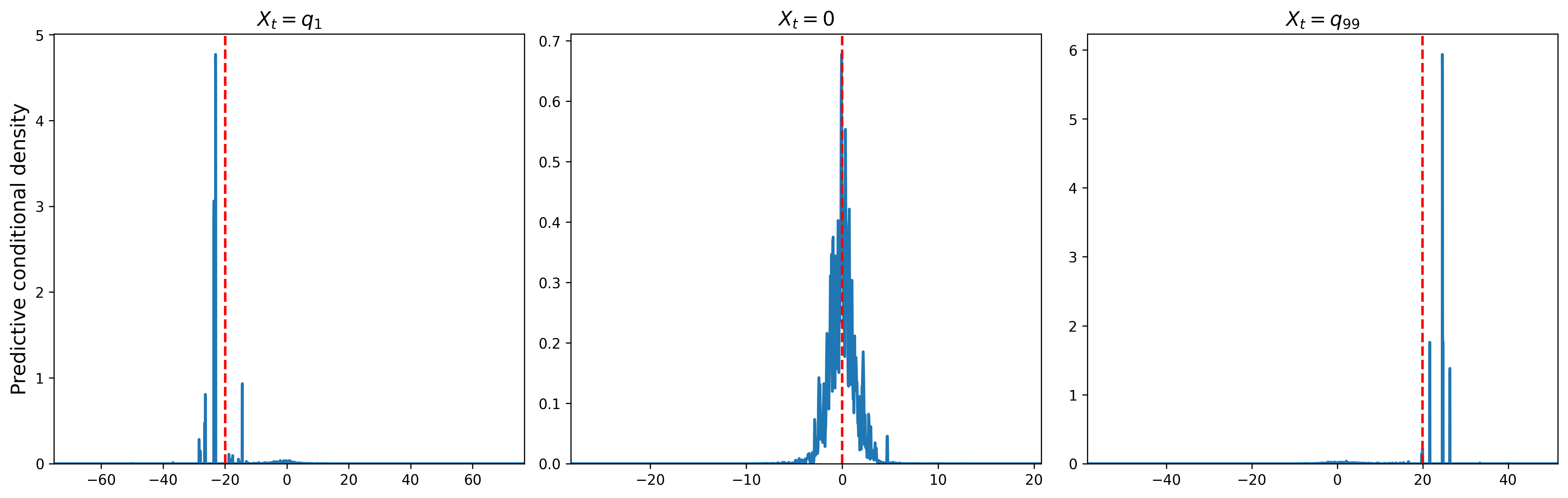}
        \caption{\cite{LanneOptimalForecasting2012}}
    \end{subfigure}
  \vspace{-0.3em}

  \begin{subfigure}{\linewidth}
        \centering
        \includegraphics[width=0.85\linewidth,height=0.14\textheight,keepaspectratio]{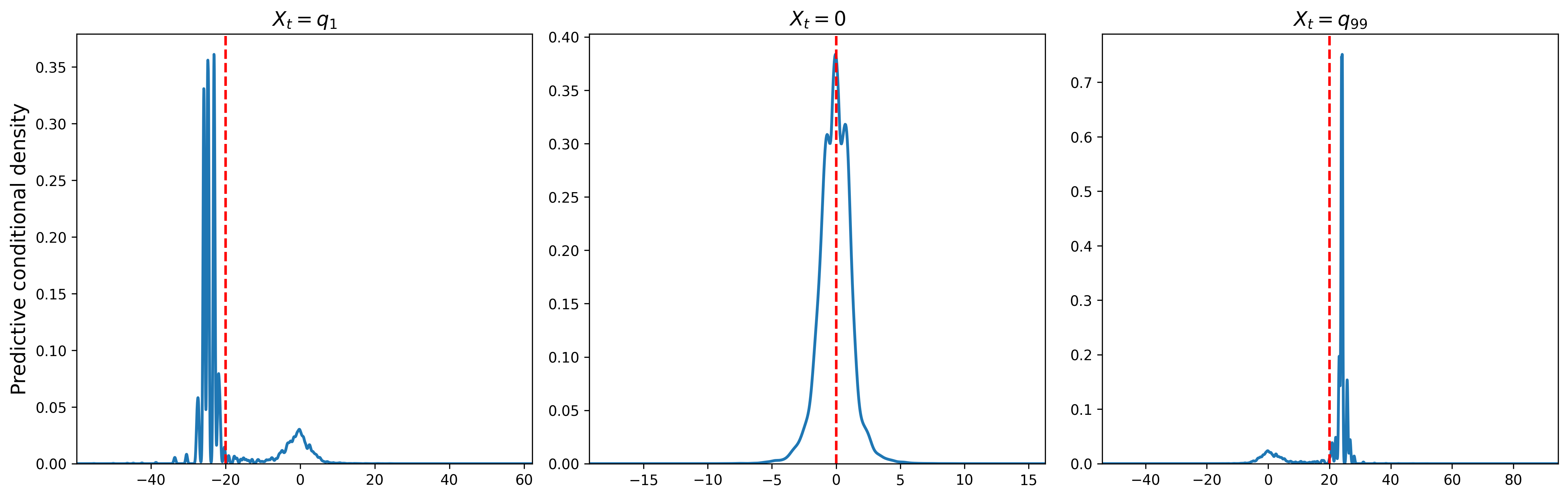}
        \caption{\cite{GourierouxNonlinearFore2026}}
    \end{subfigure}
\vspace{-0.3em}

    \begin{subfigure}{\linewidth}
        \centering
        \includegraphics[width=0.85\linewidth,height=0.14\textheight,keepaspectratio]{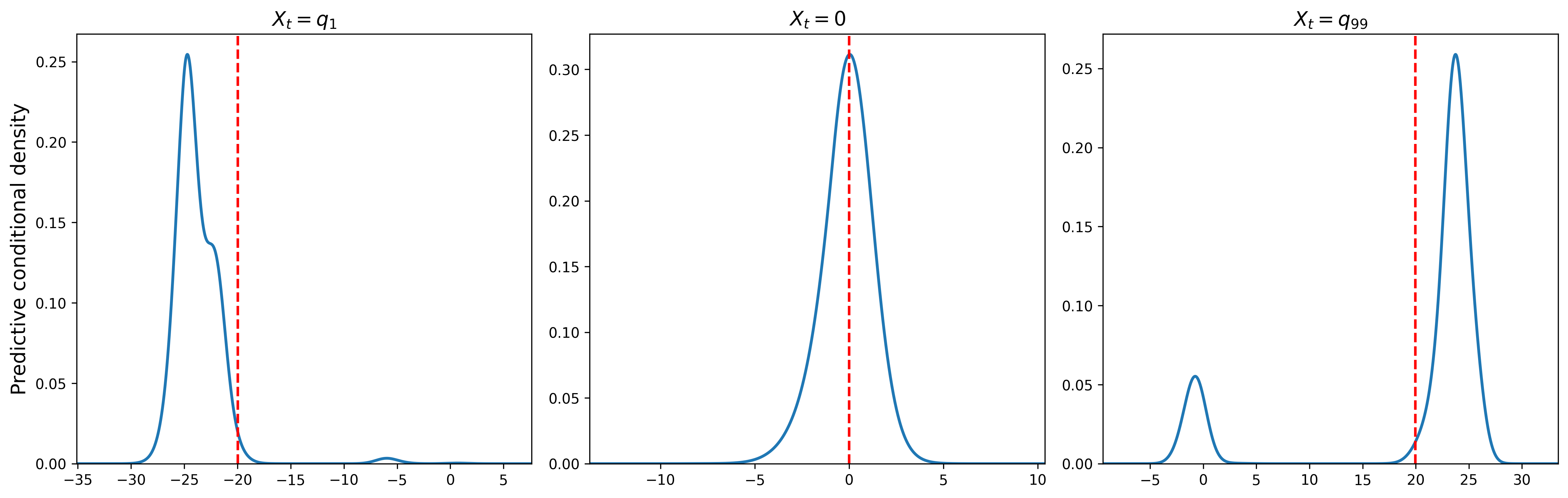}
        \caption{Nadaraya-Watson}
    \end{subfigure}
 \vspace{-0.3em}

   \begin{subfigure}{\linewidth}
        \centering
        \includegraphics[width=0.85\linewidth,height=0.14\textheight,keepaspectratio]{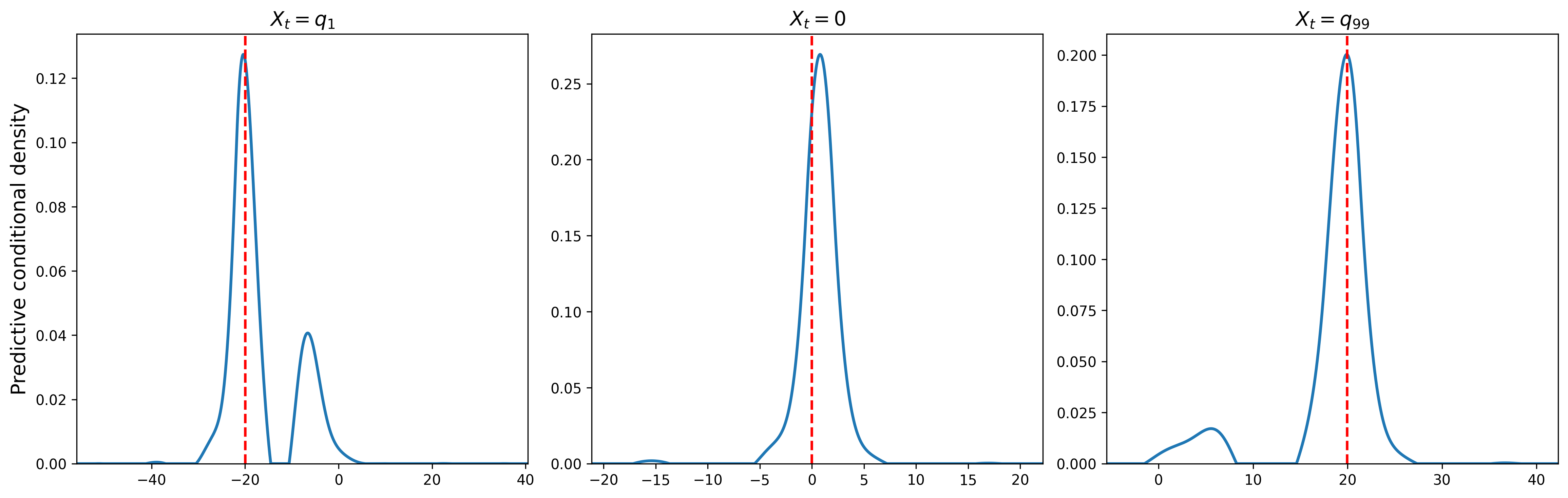}
        \caption{FlexZBoost}
    \end{subfigure}

    \caption{2-Step-Ahead Conditional Predictive Density of a MAR(0,1) Process}
    \label{fig:density1}
    \parbox{\textwidth}{\scriptsize
    \textit{Notes:} Estimated two-step-ahead predictive densities for a purely noncausal MAR(0,1) process with $\alpha$-stable innovations and $\alpha=1.4$. The three curves condition on $X_t=q_{0.01}$, $0$, and $q_{0.99}$. The red dashed line denotes the conditioning value.}
\end{figure}

Figure~\ref{fig:density1} displays the two-step-ahead conditional predictive density for a MAR(0,1) process with $\alpha=1.4$ under three conditioning scenarios: $X_t = q_{0.01}$ (left tail), $X_t = 0$ (center), and $X_t = q_{0.99}$ (right tail).
At the two-step horizon, the MDN (panel a) produces smooth predictive densities and clearly distinguishes the continuation mode near the conditioning value from the broader collapse component closer to zero in both tails.\footnote{As predicted by theory, bimodality becomes more pronounced as the horizon increases from $h=1$ to $h=5$. The results for $h=2$ are reported here; Section~3.1 of the Online Appendix presents the corresponding results for $h=1$ and $h=5$, together with animations over the full grid of 5,000 conditioning values.} This structure is less apparent for the simulation-based method of \citet{LanneOptimalForecasting2012} and the estimator of \citet{GourierouxNonlinearFore2026} (panels b and c), whose densities exhibit narrow, irregular peaks around the conditioning value and limited mass away from it. Nadaraya--Watson (panel d) recovers a secondary feature in the left tail, although its shape is not stable across tails. FlexZBoost (panel e) produces smooth densities but concentrates most of the forecast mass around the conditioning value, without clearly separating the collapse component. Overall, at $h=2$ and $\alpha=1.4$, the MDN provides the clearest and most consistent representation of the continuation-versus-collapse structure across both tails. This pattern generalizes to other values of $\alpha$ and to the other MARMA specifications with a single anticipative root.

\afterpage{\FloatBarrier}
\subsubsection{Sampled Trajectories}

\begin{figure}[!htbp]
    \centering
    \includegraphics[width=\linewidth]{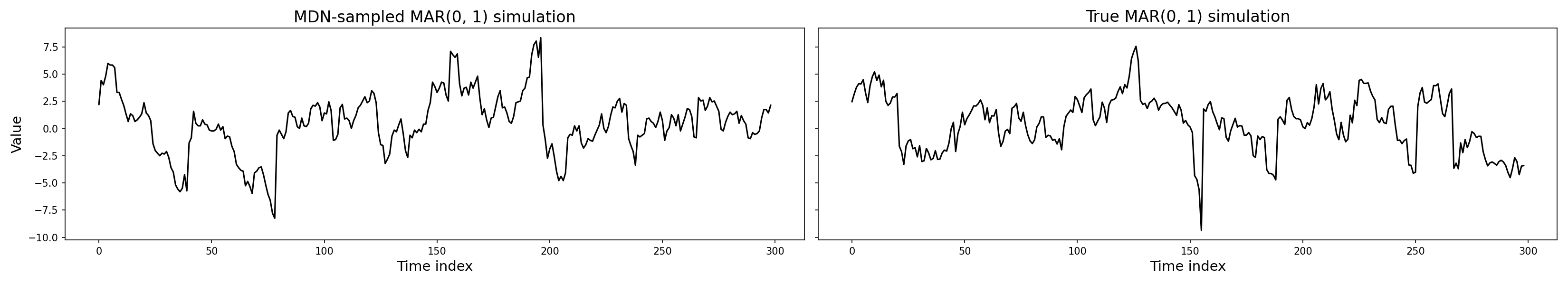}
    \caption{MDN-Sampled vs.\ True MAR(0,1) Trajectories}
    \label{fig:sampled_traj}
    \parbox{\textwidth}{\scriptsize
    \textit{Notes:} Left panel: trajectory generated by iteratively sampling from the MDN's one-step-ahead predictive density. Right panel: true MAR(0,1) simulation with $\alpha = 1.4$.}
\end{figure}

The preceding analysis examines the local shape of the conditional predictive density. Trajectory generation provides a complementary test of whether iterating the estimated one-step-ahead distribution reproduces the broader dynamics of the process. Figure~\ref{fig:sampled_traj} displays a trajectory generated by repeated sampling from the MDN's one-step-ahead predictive density.\footnote{The grid analysis in Section~\ref{sec:grid} conditions on a single value $X_t=x_i$. By contrast, trajectory generation uses the full conditioning vector $\mathbf{X}_t=(X_t,\ldots,X_{t-L+1})$ selected by the hyperparameter search described in Section~\ref{sec:overview}.} Although the MDN is trained to approximate conditional distributions rather than replicate the data-generating process directly, the sampled trajectory displays locally explosive episodes followed by abrupt reversals. This pattern indicates that iterating the estimated densities preserves the principal dynamic features associated with continuation and collapse. As a more formal assessment, we estimate a MAR(0,1) model on the MDN-generated trajectory. Table~\ref{tab:mar01_sampled} shows that the resulting parameter estimates are close to those of the original data-generating process. This exercise provides complementary evidence that the MDN captures key statistical features of the underlying noncausal dynamics.

Section 3.1 of the Online Appendix reports the same exercise for the MAR(0,2), MAR(1,1) and MARMA(1,1,1,1) specifications, showing that the approach extends beyond the benchmark MAR(0,1) case.

\begin{table}[!htbp]
\centering
\caption{Estimated MAR(0,1) parameters from the MDN-sampled trajectory ($\alpha = 1.4$)}
\label{tab:mar01_sampled}
\begin{threeparttable}
\begin{tabular}{lcc}
\toprule
Parameter & True & Estimates (ML) \\
\midrule
$\psi$ & 0.900 & \makecell{0.905$^{***}$ \\ (0.017)} \\
$\alpha$ & 1.400 & \makecell{1.593$^{***}$ \\ (0.084)} \\
$\beta$ & 0.000 & \makecell{0.107 \\ (0.177)} \\
$\sigma$ & 0.500 & \makecell{0.525$^{***}$ \\ (0.031)} \\
\bottomrule
\end{tabular}
\begin{tablenotes}[para,flushleft]
\footnotesize
\item \textit{Note:} Standard errors in parentheses. $^{***}$ $p<0.01$, $^{**}$ $p<0.05$, $^{*}$ $p<0.10$.
\end{tablenotes}
\end{threeparttable}
\end{table}

\afterpage{\FloatBarrier}
\subsection{Simulation Results}

\subsubsection{Overview}
\label{sec:overview}

The objective of the simulation study is to assess the accuracy with which each method recovers the predictive density implied by anticipative processes. Since the true conditional density is unavailable for most MARMA specifications, we conduct three complementary exercises, each comparing the estimated densities with a different external benchmark:

\begin{enumerate}
\item[Case 1] \emph{Against the true density.} For the Cauchy MAR(0,1) process ($\alpha = 1.0$, $\beta = 0$), the conditional predictive density $p_h$ is available in closed form \citep{GourierouxLocalExplosion2017}. Each estimate $\hat{p}_h$ is compared to it through the Kullback-Leibler divergence, and the integrated squared error.
\item[Case 2] \emph{Against realized outcomes.} When $p_h$ is unavailable, each $\hat{p}_h(\cdot \mid \mathbf{X}_t)$ is scored against the realization $X_{t+h}$ of an independently simulated path, through proper scoring rules $S(\hat{p}_h, X_{t+h})$ averaged over the evaluation sample.
\item[Case 3] \emph{Against theoretical conditional moments.} For general MARMA processes, \citet{FriesConditionalMoments2022} provide closed-form expressions for the conditional moments whenever the relevant existence conditions are satisfied. For each $\alpha$, we evaluate only the integer orders $m \in \{1,\ldots,4\}$ satisfying $m < 2\alpha+1$. The estimated moments \[ \hat{\mathbb E}[X_{t+h}^{m}\mid X_t=x] =\int u^{m}\hat p_h(u\mid x)\,du \] are compared with their theoretical counterparts using the root mean squared error.
\end{enumerate} 

The three exercises move from the strongest benchmark available (true density) to settings where only partial theoretical information is available.
All exercises consider horizons $h\in\{1,2,5\}$. In Cases 1 and 3, performance is evaluated over a fixed grid of conditioning values spanning $[q_{0.01},q_{0.99}]$.\footnote{The grid is univariate because the closed-form predictive density of \citet{GourierouxLocalExplosion2017} and the theoretical conditional moments of \citet{FriesConditionalMoments2022} are defined for univariate conditioning, i.e. conditional on $X_t$ alone rather than on the full vector $\mathbf{X}_t$.} We report results for the center, $[q_{0.10},q_{0.90}]$; the tails, $[q_{0.01},q_{0.10}]\cup[q_{0.90},q_{0.99}]$; and the full grid. Case 2 evaluates density forecasts against realized outcomes. \emph{Ex post} selection of observations according to whether the realization is extreme would render an otherwise proper scoring rule improper \citep{LerchEtAl2017}. We therefore evaluate conventional scores over the full sample and assess tail performance using the threshold-weighted CRPS \citep{GneitingRanjan2011} and censored likelihood score \citep{DiksEtAl2011}. These scores weight regions of the predictive support rather than selecting observations according to realized outcomes.

Each comparison is accompanied by the 75\% Model Confidence Set (MCS) of \citet{HansenConfidenceSet2011}, which retains methods that are statistically indistinguishable from the best-performing method. In the tables, an asterisk denotes MCS membership.

The theoretical benchmarks used in Cases 1 and 3 condition only on $X_t$ \citep{GourierouxLocalExplosion2017,FriesConditionalMoments2022}. To ensure comparability, we therefore set $L=1$ for all methods in these exercises. The model-based methods of \citet{LanneOptimalForecasting2012} and \citet{GourierouxNonlinearFore2026}, however, require $p+q$ lags to filter the noncausal state. They can consequently be evaluated in Case 3 only for MAR(0,1), where $p+q=L=1$. MAR(0,2) and MAR(1,1) require two lags, while MARMA(1,1,1,1) lies outside the class of models accommodated by these methods.

Hyperparameters are selected using the 1,000-observation calibration sample. For the MDN, Bayesian optimization with a Tree-structured Parzen Estimator over 100 trials selects the depth $D$, width $H$, number of mixture components $K$, dropout rate, and learning rate \citep{AkibaOptunaNext2019}. In Case 2, the number of conditioning lags $L$ is also selected. Each trial trains a candidate network on the 2,000-observation training trajectory, and the configuration achieving the lowest calibration-sample negative log-likelihood is retained for forecasting. For Nadaraya--Watson and FlexZBoost, only $L$ is tuned in Case 2, using the log-likelihood and CDE loss, respectively. Table~\ref{tab:hyperparams} reports the corresponding search spaces.

\begin{table}[!htbp]
\centering
\small
\caption{Hyperparameter Search Space}
\label{tab:hyperparams}
\begin{threeparttable}
\begin{tabular}{ll}
\toprule
Hyperparameter & Search space \\
\midrule
\multicolumn{2}{l}{\textit{Mixture Density Network} (Tree-structured Parzen Estimator, 100 trials)} \\
\quad Depth $D$ & $\{1, \ldots, 5\}$ \\
\quad Width $H$ & $\{64, 128, 256, 512\}$ \\
\quad Mixture components $K$ & $\{1, \ldots, 10\}$ \\
\quad Dropout rate & $\{0, 0.05, 0.10, 0.15, 0.20\}$ \\
\quad Learning rate & \makecell[l]{$\{0.0001, 0.0002, 0.0004, 0.0006, 0.0008,$\\$\phantom{\{}0.001, 0.002, 0.004, 0.006, 0.008, 0.01\}$} \\
\quad Lags $L$ & $\{1, \ldots, 10\}$ \\
\midrule
\multicolumn{2}{l}{\textit{Nadaraya-Watson}} \\
\quad Kernel & Gaussian (fixed) \\
\quad Bandwidth & Silverman's rule (fixed) \\
\quad Lags $L$ & $\{1, \ldots, 10\}$ \\
\midrule
\multicolumn{2}{l}{\textit{FlexZBoost}} \\
\quad Basis system & Cosine (fixed) \\
\quad Maximum number of basis terms & 31 (fixed) \\
\quad Lags $L$ & $\{1, \ldots, 10\}$ \\
\bottomrule
\end{tabular}
\begin{tablenotes}[para,flushleft]
\footnotesize
\item \textit{Note:} Hyperparameters are selected using the 1,000-observation calibration sample. In Cases 1 and 3, all methods condition only on $X_t$ ($L=1$). The methods of \citet{LanneOptimalForecasting2012} and \citet{GourierouxNonlinearFore2026} have no tunable hyperparameters and use the $p+q$ lags required for state filtering. 
\end{tablenotes}
\end{threeparttable}
\end{table}

\subsubsection{Case 1: Cauchy MAR(0,1) model}

The first exercise turns to the Cauchy MAR(0,1) process ($\alpha = 1.0$, $\beta = 0$), for which the predictive density is available in closed form. Table \ref{tab:density_mar01} compares the estimated predictive densities with this true density, using Kullback-Leibler (KL) divergence and Integrated Squared Error (ISE) metrics.\footnote{The KL divergence is defined as $\text{KL}(p_h \| \hat{p}_h) = \int p_h(X_{t+h} \mid \mathbf{X}_t) \log \left( \frac{p_h(X_{t+h} \mid \mathbf{X}_t)}{\hat{p}_h(X_{t+h} \mid \mathbf{X}_t)} \right) dX_{t+h}$, while the ISE is given by $\text{ISE} = \int \left( p_h(X_{t+h} \mid \mathbf{X}_t) - \hat{p}_h(X_{t+h} \mid \mathbf{X}_t) \right)^2 dX_{t+h}$.}
\ExecuteMetaData[outputs/simulations.tex]{density-mar01}

The MDN consistently attains the lowest KL divergence and ISE in the tails and over the full distribution at every forecast horizon, and it is the only method belonging to the MCS in these two regions. The margin is wide for the KL divergence, which penalizes the regions where a method assigns too little mass. At $h=5$, the tail KL divergence is 0.777 for the MDN, against 3.264 for FlexZBoost and 7.627 for Nadaraya-Watson. In the center, the estimator of \cite{GourierouxNonlinearFore2026} performs best at $h=1$ and $h=2$, but degrades at $h=5$, as it only predicts one-step-ahead and must chain its own forecasts to reach longer horizons. Across all horizons, regions and metrics, the MDN ranks first or second in every cell. It is first in fourteen of the eighteen cells and second in the four remaining ones. All four occur in the center at $h=1$ and $h=2$, where the estimator of \cite{GourierouxNonlinearFore2026} leads with differences as small as 0.001 in ISE at $h=2$. Each competitor is thus accurate in one region at the cost of the other, whereas the MDN is the only method that remains accurate in the center and in the tails simultaneously.

\subsubsection{Case 2: Comparison with the Realized Outcomes}
\label{sec:case2}

Having examined density recovery in a setting where the predictive density is known analytically, we now evaluate the forecasts against realized outcomes 
 $X_{t+h}$. Each fitted method produces $h$-step-ahead forecasts along an independently simulated trajectory of 1,000 observations. We score the resulting densities using six complementary scoring criteria: the Conditional Density Estimation (CDE) loss, the Continuous Ranked Probability Score (CRPS), the logarithmic probability score, the quantile scores at the 5\%, 10\%, 90\% and 95\% levels, the threshold-weighted CRPS, and the censored likelihood score at the same four thresholds. Section 3.2 of the Online Appendix gives the formal definitions.\footnote{For CDE loss, CRPS, and quantile scores, lower values indicate superior performance, while for the log probability score, higher values are preferred.}

Tables~\ref{tab:mar01_metrics_mar01_h1}, \ref{tab:mar01_metrics_mar02_h1}, \ref{tab:mar01_metrics_mar11_h1} and \ref{tab:mar01_metrics_marma1111_h1} report the one-step-ahead results. For each of the four processes, the MDN records the best value in between 70 and 75 of the 75 metric-by-tail-index cells, 294 of the 300 cells in total. It sweeps MAR(0,1) and MAR(0,2), and cedes a single cell on MAR(1,1) to \cite{GourierouxNonlinearFore2026}. The six cells it loses are concentrated on the MARMA specification at $\alpha = 1.6$ and $\alpha = 1.8$, where FlexZBoost takes the CDE loss and the censored likelihood score at the 5\% and 10\% levels; the MDN is second in five of the six. Results for $h = 2$ and $h = 5$ are collected in Section 3.2 of the Online Appendix; the rankings are essentially unchanged.
\ExecuteMetaData[outputs/simulations_realized_outcomes.tex]{mar01-metrics-mar01-h1}

\ExecuteMetaData[outputs/simulations_realized_outcomes.tex]{mar01-metrics-mar02-h1}

\ExecuteMetaData[outputs/simulations_realized_outcomes.tex]{mar01-metrics-mar11-h1}

\ExecuteMetaData[outputs/simulations_realized_outcomes.tex]{mar01-metrics-marma1111-h1}

\afterpage{\FloatBarrier}
\subsubsection{Case 3: Predictive Moments Approach}

The final exercise compares conditional moments obtained from the estimated densities with their theoretical counterparts. For each value of $\alpha$, we consider the integer orders $m\leq4$ satisfying $m<2\alpha+1$. Moment accuracy is measured by RMSE over the evaluation grid.
Tables~\ref{tab:rmse_mar01_h1}, \ref{tab:rmse_mar02_h1}, \ref{tab:rmse_mar11_h1} and \ref{tab:rmse_marma1111_h1} report the results.

The MDN attains the lowest RMSE in 40 of the 64 tail cells and the same number over the full distribution. It is also the most frequent winner in the center, leading in 46 of 64 cells, compared with 9 for \citet{GourierouxNonlinearFore2026} and 7 for Nadaraya--Watson. Its advantage is most pronounced for MAR(0,2) and MAR(1,1), where it leads in 12 and 16 tail cells, respectively, across moment orders.
The main exceptions concern MAR(0,1) and MARMA(1,1,1,1). For MAR(0,1), the estimator of \citet{GourierouxNonlinearFore2026} generally performs best in the tails for $\alpha\geq1.4$. For MARMA(1,1,1,1), FlexZBoost leads for the first moment up to $\alpha=1.6$, while Nadaraya--Watson leads at $\alpha=1.8$.\footnote{These exceptions arise primarily as $\alpha$ increases and the tails become lighter. Nadaraya--Watson also leads across the MAR(0,2) tail cells at $\alpha=1.8$, but not at lower values of $\alpha$ or for MAR(1,1).} The method of \citet{LanneOptimalForecasting2012} is competitive in only two center cells, while FlexZBoost records no wins outside the MARMA tails.

\ExecuteMetaData[outputs/simulations.tex]{rmse-mar01-h1}

\ExecuteMetaData[outputs/simulations.tex]{rmse-mar02-h1}

\ExecuteMetaData[outputs/simulations.tex]{rmse-mar11-h1}

\ExecuteMetaData[outputs/simulations.tex]{rmse-marma1111-h1}

\afterpage{\FloatBarrier}

The MCS results reinforce these rankings. A single method remains in the $MCS_{75\%}$ in 55 of the 64 tail cells and 53 of the 64 full-distribution cells; in each region, the MDN is that method in 39 cases.\footnotemark\pagebreak[3]
\footnotetext{Tables S12 to S19 of Section 3.3 in the Online Appendix repeat the exercise at $h = 2$ and $h = 5$; the conclusions are qualitatively unchanged.}

\begin{figure}[!htbp]
    \centering
    \begin{subfigure}[b]{0.46\linewidth}
        \centering
        \includegraphics[width=\linewidth]{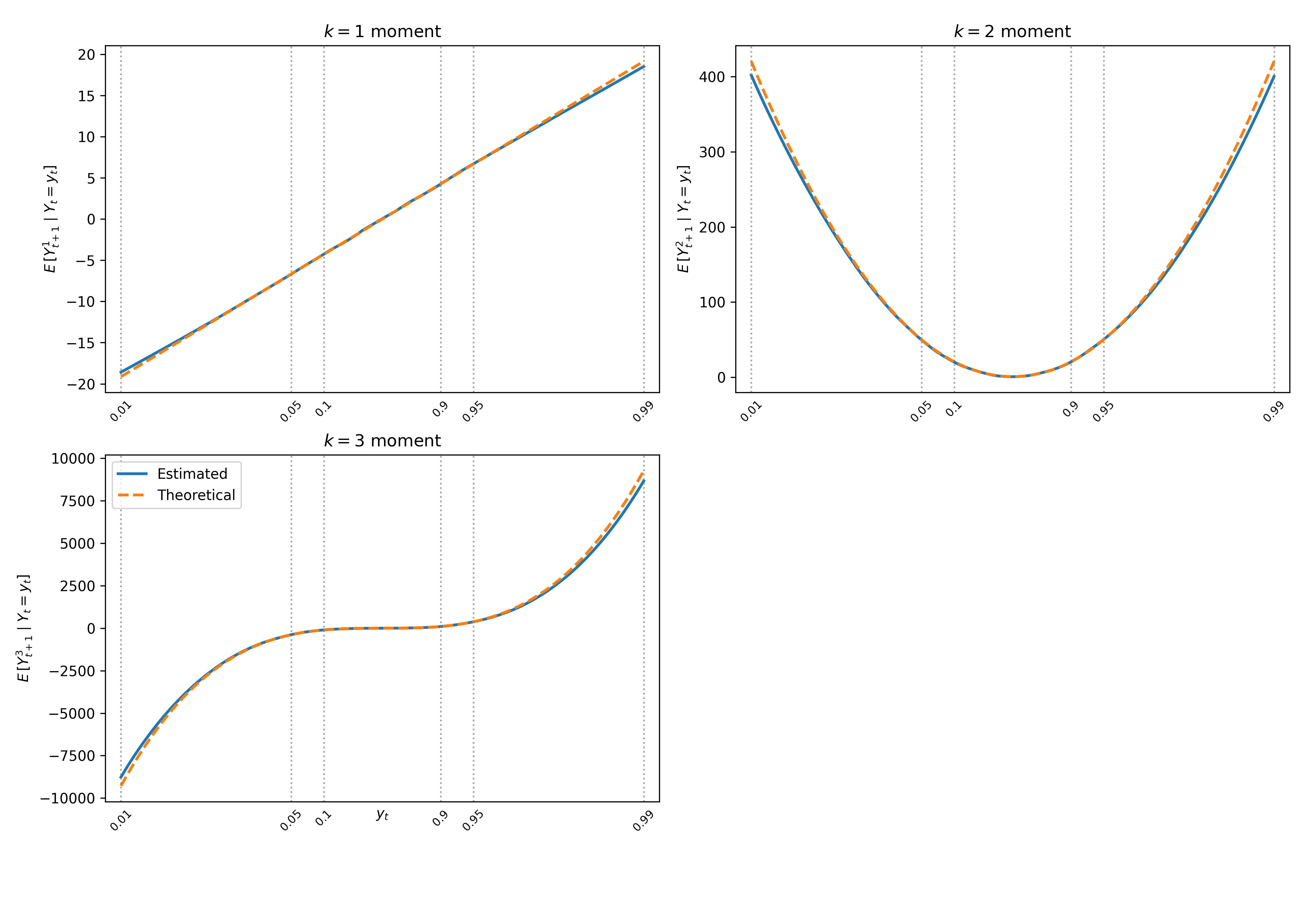}
        \caption{Mixture Density Network}
    \end{subfigure}

    \vspace{0.3em}

    \begin{subfigure}[b]{0.46\linewidth}
        \centering
        \includegraphics[width=\linewidth]{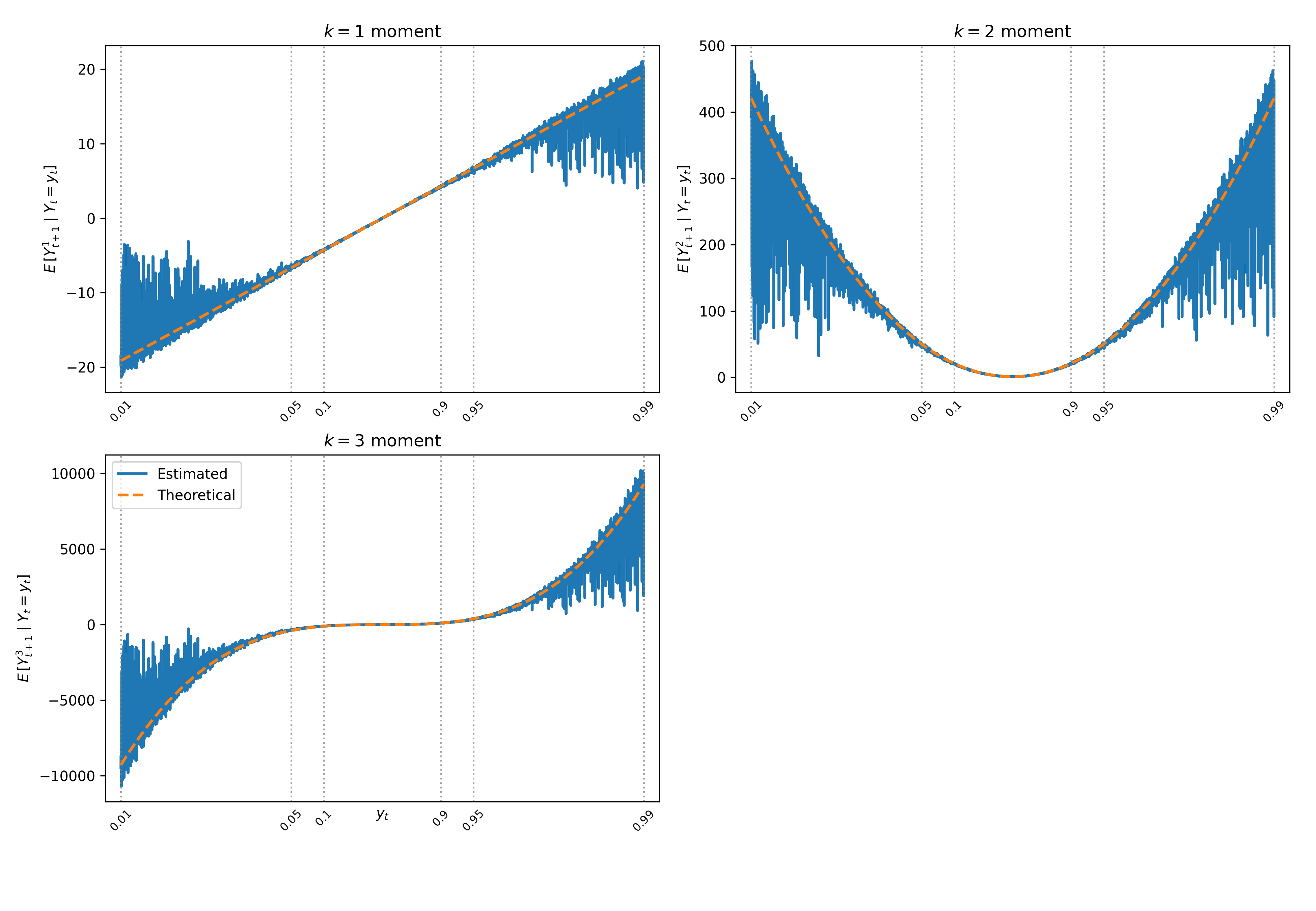}
        \caption{Lanne--Saikkonen}
    \end{subfigure}
    \hfill
    \begin{subfigure}[b]{0.46\linewidth}
        \centering
        \includegraphics[width=\linewidth]{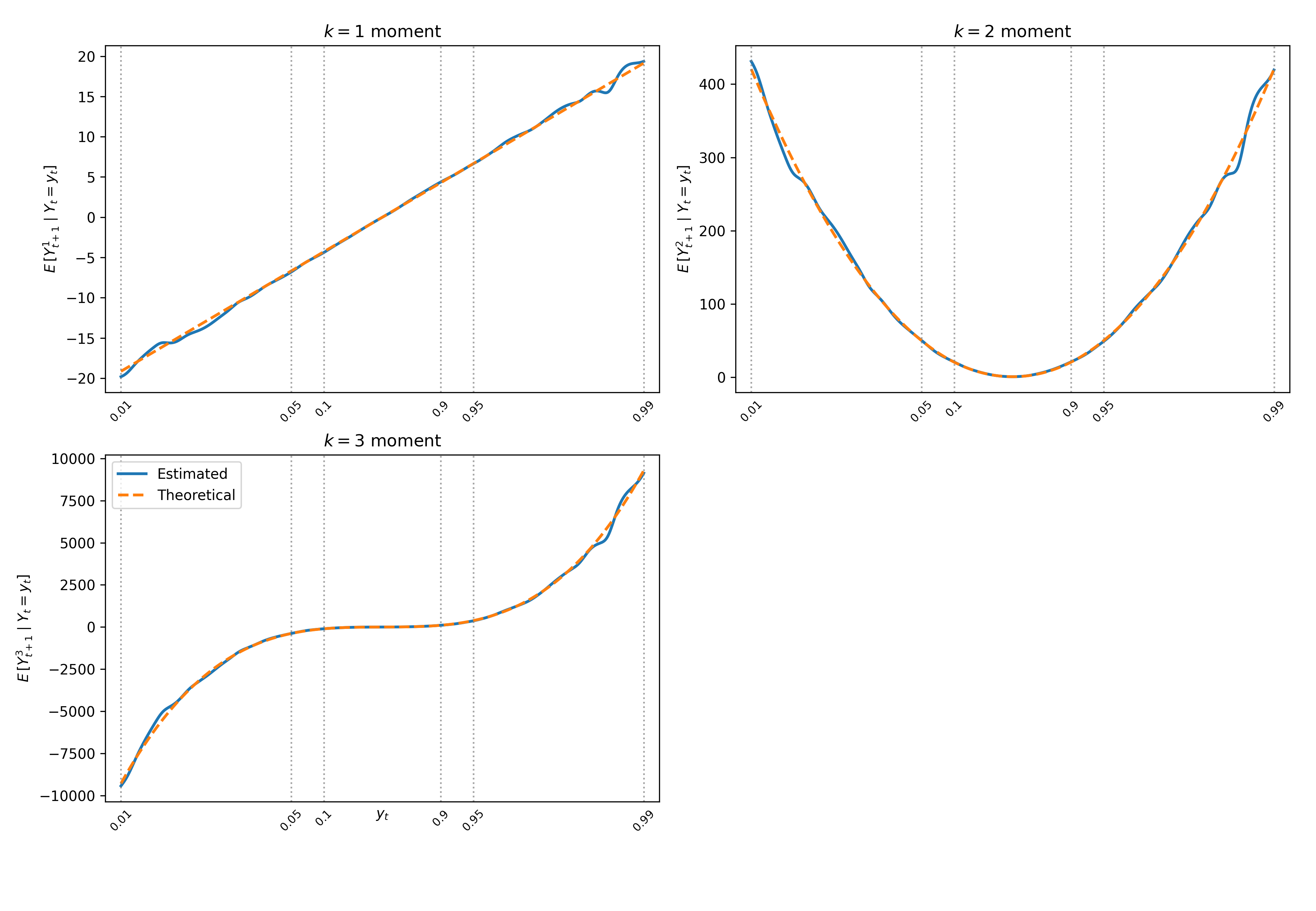}
        \caption{Gouri\'eroux--Jasiak}
    \end{subfigure}

    \vspace{0.3em}

    \begin{subfigure}[b]{0.46\linewidth}
        \centering
        \includegraphics[width=\linewidth]{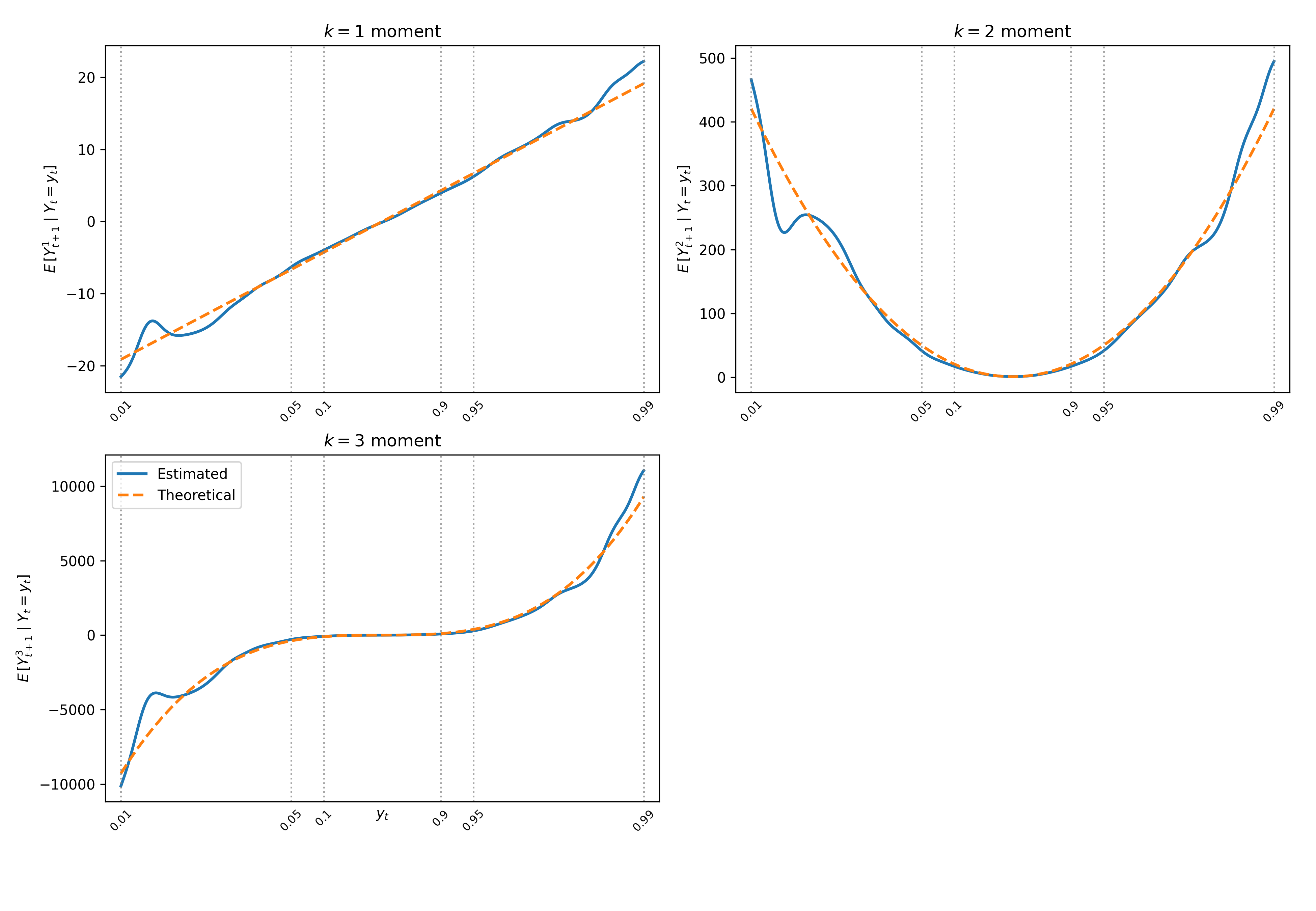}
        \caption{Nadaraya-Watson}
    \end{subfigure}
    \hfill
    \begin{subfigure}[b]{0.46\linewidth}
        \centering
        \includegraphics[width=\linewidth]{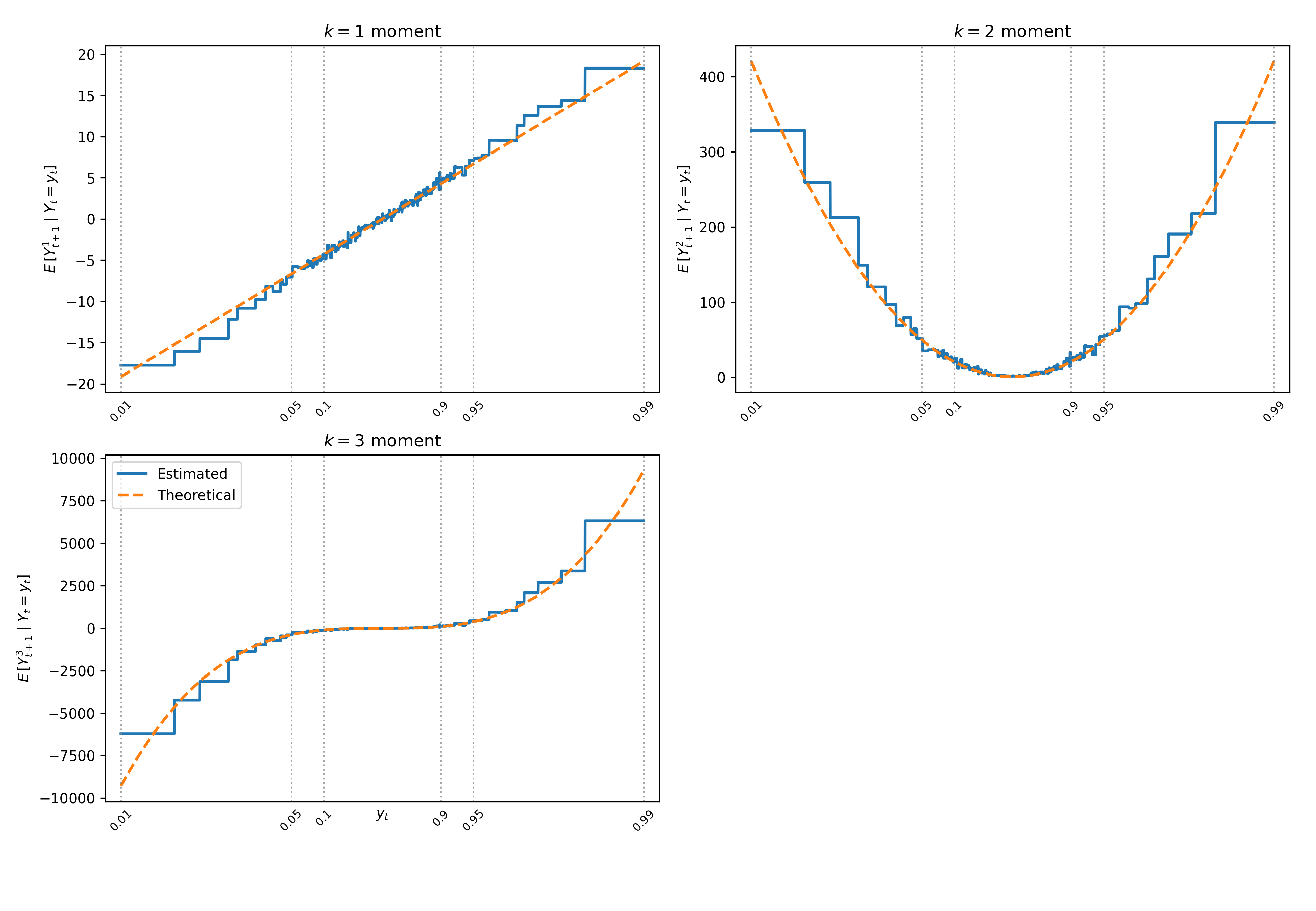}
        \caption{FlexZBoost}
    \end{subfigure}

  \caption{ Conditional Predictive Moment Accuracy: MAR(0,1) Process, 1-Step-Ahead Forecasts}
    \label{f1}
    \parbox{\textwidth}{\scriptsize
    \textit{Notes:} Estimated and theoretical conditional moments $\mathbb{E}[X_{t+h}^m\mid X_t]$, for $m\in\{1,2,3\}$, of a purely noncausal MAR(0,1) process with $\alpha$-stable innovations and $\alpha=1.4$. Blue curves denote estimated moments; orange curves denote their theoretical counterparts. Panels (a)--(e) correspond to the five forecasting methods.  }
\end{figure}

Figure~\ref{f1} compares the one-step-ahead conditional moments for a MAR(0,1) process with tail index $\alpha=1.4$. The MDN produces smooth estimates that closely track the theoretical moments over the evaluation range. The estimates of \citet{GourierouxNonlinearFore2026} and Nadaraya--Watson are also smooth, but their deviations from the theoretical curves increase away from the center. FlexZBoost exhibits a staircase pattern and tends to understate the moments in the tails. The simulation-based method of \citet{LanneOptimalForecasting2012} displays the greatest variability, particularly away from the center. Similar, though noisier, patterns arise at horizons $h=2$ and $h=5$; the corresponding results are reported in the Online Appendix.

Moment comparisons nevertheless provide a less demanding assessment than density-based criteria because integration can offset local approximation errors. Accordingly, the competing methods appear closer to the MDN in this exercise than under the density-based evaluations in Cases 1 and 2.
\FloatBarrier

\subsubsection{{Summary of the Simulation Evidence}}

The three exercises use different benchmarks but yield broadly consistent rankings. Against the closed-form Cauchy density, the MDN attains the lowest KL divergence and ISE in the tails and over the full range at every horizon. Against realized outcomes, it records the best score in 294 of the 300 cells and ranks second in five of the remaining six. In the conditional-moment comparison, the MDN is the most frequent winner in every region, although the estimator of \citet{GourierouxNonlinearFore2026} leads in a substantial subset of the MAR(0,1) cases.

Overall, the MDN provides the most accurate approximation to the noncausal predictive densities among the methods considered. Its advantage is most pronounced in the tails, where the continuation-versus-collapse structure is strongest.


\section{Empirical Application}
\label{s5}

This section presents a real-time application to U.S. natural gas prices. The predictive densities of the MDN are compared with those of the competing estimators, and with the benchmarks of the energy forecasting literature.

\subsection{Forecasting Natural Gas Prices in Real Time}
\label{s5a}

Having evaluated the MDN in controlled environments where the data-generating process is known, we now turn to a real-time forecasting exercise in which both model uncertainty and data revisions are present. Natural gas prices provide a particularly relevant testing ground because they exhibit recurrent episodes of locally explosive behaviour and abrupt reversals.

We adopt the real-time forecasting framework of \cite{BaumeisterForecastingNatural2026}, who provide an extensive evaluation of point forecasting methods for the real Henry Hub spot price, the benchmark for North American natural gas markets. Their setup accounts for publication lags and data revisions through a database of monthly vintages, ensuring that forecasts rely only on information available at each forecast origin. The estimation period runs from January 1976 to January 1997, and the out-of-sample evaluation period spans February 1997 to February 2024. We defer the reader to \cite{BaumeisterForecastingNatural2026} for a detailed description of the data construction and real-time constraints.
\begin{figure}[!htbp]
\centering
\includegraphics[width=0.65\textwidth]{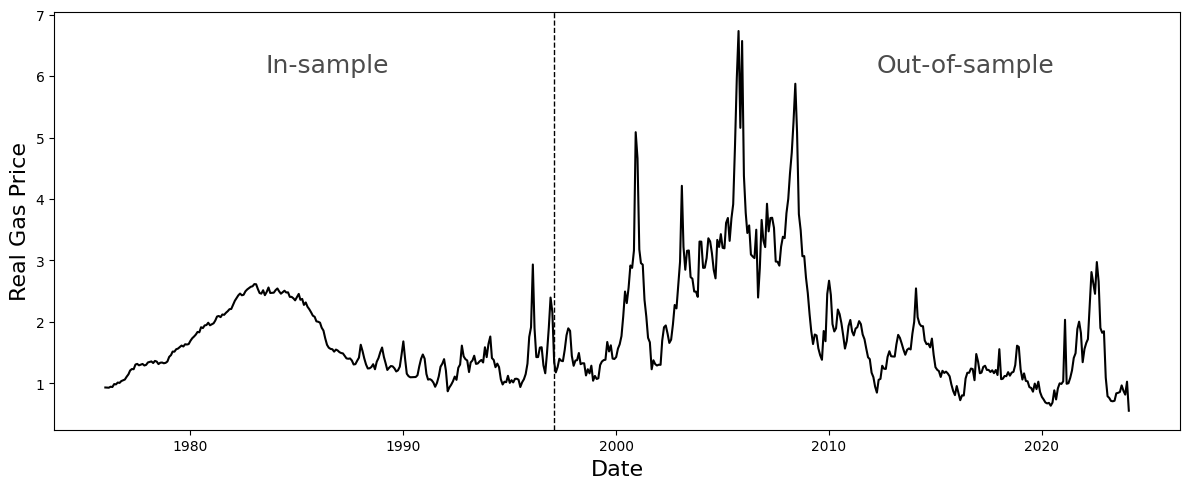}
\caption{Real Henry Hub spot price of natural gas. 
}
\label{fig:gas_price}
\end{figure}

Figure~\ref{fig:gas_price} displays the real Henry Hub spot price over the sample period. The series exhibits several episodes of rapid appreciation followed by sharp reversals, most notably during 2005--2008, consistent with the dynamics that noncausal models are designed to capture.

Following \citet{BaumeisterForecastingNatural2026}, we model the logarithm of the price and transform the forecasts back to levels. Point forecasts are obtained by exponentiation. For density forecasts, we additionally apply the Jacobian of the transformation: if $P_{t+h}=\exp(X_{t+h})$, the predictive density of the price is \[ \hat{p}^{P}_h(P_{t+h}\mid\mathbf{X}_t) = \frac{\hat{p}_h(X_{t+h}\mid\mathbf{X}_t)}{P_{t+h}}. \] We apply the two-stage estimation procedure described in Section~\ref{sec:param-estimation} to log prices over the initial estimation period, 1976M1--1997M1. The retained specification is a mixed causal--noncausal MAR(1,1), selected as the most parsimonious model whose residuals pass the GCov portmanteau specification test of \citet{JasiakGcovPortmanteau2026}.

Table~\ref{tab:gas_params} reports both the maximum-likelihood and the GCov estimates.
This specification is robust over time: the full post-revised series (1976M1--2024M2) delivers comparable estimates. Both samples show strong anticipative persistence and heavy tails, with a tail index well below the Gaussian value $\alpha = 2$, consistent with the large price movements visible in Figure~\ref{fig:gas_price}.

\begin{table}[!htbp]
\centering
\caption{Estimated MAR(1,1) parameters for the real Henry Hub spot price}
\label{tab:gas_params}
\begin{threeparttable}
\begin{tabular}{lcccc}
\toprule
 & \multicolumn{2}{c}{Real-Time In-Sample} & \multicolumn{2}{c}{Full Period Post-Revised} \\
\cmidrule(lr){2-3} \cmidrule(lr){4-5}
Parameter & ML & GCov & ML & GCov \\
\midrule
$\phi_{1}$ & \makecell{0.977$^{***}$ \\ (0.010)} & \makecell{0.865$^{***}$ \\ (0.021)} & \makecell{0.969$^{***}$ \\ (0.010)} & \makecell{0.965$^{***}$ \\ (0.020)} \\
 & & & & \\
$\psi_{1}$ & \makecell{0.242$^{***}$ \\ (0.047)} & \makecell{0.301$^{***}$ \\ (0.078)} & \makecell{0.118$^{***}$ \\ (0.035)} & \makecell{0.021 \\ (0.043)} \\
 & & & & \\
$\alpha$ & \makecell{1.155$^{***}$ \\ (0.078)} & \makecell{1.641$^{***}$ \\ (0.179)} & \makecell{1.350$^{***}$ \\ (0.079)} & \makecell{1.686$^{***}$ \\ (0.120)} \\
 & & & & \\
$\beta$ & \makecell{$-0.052$ \\ (0.125)} & \makecell{$-0.020$ \\ (0.179)} & \makecell{0.016 \\ (0.086)} & \makecell{0.045 \\ (0.118)} \\
 & & & & \\
$\sigma$ & \makecell{0.023$^{***}$ \\ (0.002)} & \makecell{0.037$^{***}$ \\ (0.003)} & \makecell{0.053$^{***}$ \\ (0.003)} & \makecell{0.060$^{***}$ \\ (0.004)} \\
\bottomrule
\end{tabular}
\begin{tablenotes}[para,flushleft]
\footnotesize
\item \textit{Note:} Standard errors in parentheses. $^{***}$ $p<0.01$, $^{**}$ $p<0.05$, $^{*}$ $p<0.10$.
\end{tablenotes}
\end{threeparttable}
\end{table}

Following \cite{BaumeisterForecastingNatural2026}, we re-estimate the MAR(1,1) parameters at each forecast origin, on the vintage available at that date. They are subsequently used to generate a new training trajectory of 2,000 observations and a calibration sample of 1,000 observations. All forecasting methods are then re-estimated on these simulated data before producing the forecast for that origin. Hyperparameters are tuned at the first forecast origin only. The architectures are then held fixed, and only the estimated parameters change from one origin to the next.

Several features of the training procedure and of the evaluation protocol are adapted to the empirical application. First, whereas the simulations treat both tails symmetrically, the real gas price is bounded below by zero, so that the locally explosive trajectories we aim to forecast are concentrated in the upper tail. We therefore restrict the tail-state set defined in Section~\ref{s3} to its upper branch, \[ \mathcal{E}=\{t\in\mathcal{D}:X_t>u\}, \] while leaving the weighting scheme otherwise unchanged.

Second, the tail region over which the models are evaluated also differs. Rather than by quantiles of a theoretical marginal distribution, it is delimited by the 90th percentile of the real prices available in the origin's own vintage, and is therefore known when the forecast is made. A forecast origin is a tail origin when its current observed price exceeds this threshold. The classification thus precedes the realization of the forecast outcome. At the one-month horizon, 73 of the 325 forecast origins satisfy this criterion.

Finally, the simulations report the weighted scores and the quantile score at the four levels $\tau\in\{0.05,0.10,0.90,0.95\}$. Our empirical application uses the same metrics but only at the upper levels, for the reason just given. The threshold-weighted CRPS and the censored likelihood score are restricted further to $\tau=0.90$: the 325 forecast origins leave too few extreme realizations to evaluate them reliably at $\tau=0.95$. The quantile score is reported at both $\tau\in\{0.90,0.95\}$.

We generate out-of-sample forecasts over the period 1997M2--2024M2 for horizons $h \in \{1, 3, 6, 9, 12, 15, 18, 21, 24\}$ months.

\afterpage{\FloatBarrier}
\subsection{Forecasting Results}


The empirical evaluation proceeds in three steps, each confronting the MDN with a different set of competitors:
\begin{enumerate}
\item[Level 1] \emph{Which density estimator?} The five estimators of Section~\ref{SD} are all fitted on the same estimated MAR(1,1), so that the comparison isolates the density estimator itself.
\item[Level 2]  \emph{Does the noncausal specification pay?} The MDN is confronted with the closed-form Gaussian predictive densities of the three best causal linear benchmarks of \cite{BaumeisterForecastingNatural2026}, so that the comparison isolates the specification rather than the estimator.
\item[Level 3]  \emph{Do the density gains survive in point forecasting?} The median of the MDN predictive density is compared with the univariate and BVAR benchmarks of \cite{BaumeisterForecastingNatural2026} in terms of mean squared prediction error.
\end{enumerate}

\subsubsection{Density Forecasts: Comparing Estimators on the Same Causal--Noncausal Model}

This first level holds the model fixed and varies the estimator. The five approaches of Section~\ref{SD} are all fitted on trajectories simulated from the same estimated MAR(1,1), so that any difference in performance is imputable to the density estimator alone.

Table~\ref{tab:gas_comparison} summarizes the results. The MDN ranks first in 48 of the 63 metric-by-horizon cells and second in the remaining 15, while belonging to the $MCS_{75\%}$ in 61 cells. Nadaraya--Watson is the closest competitor, ranking first in 14 cells, primarily at $h=6$ and $h=9$, and second in 37. FlexZBoost and the methods of \citet{LanneOptimalForecasting2012} and \citet{GourierouxNonlinearFore2026} collectively lead in one cell. The weighted scores indicate that the MDN's advantage is particularly pronounced in the upper tail. It attains the lowest censored likelihood score at all nine horizons, outperforming Nadaraya--Watson by between 2.0\% and 12.1\%, and the lowest threshold-weighted CRPS at six horizons. Nadaraya--Watson leads at $h=6$ and $h=9$, while FlexZBoost leads at $h=1$.

These findings are consistent with the Monte Carlo evidence, where the MDN's advantage was most pronounced in the tails. Although the weighted scores use all forecast origins, only 74 of the 325 realized prices lie in the emphasized upper-tail region. Upper-tail performance is therefore assessed with substantially less information than in the simulations. Despite this difference, the real-time results broadly support the simulation ranking.

\ExecuteMetaData[outputs/applications.tex]{gas-comparison}

\afterpage{\FloatBarrier}
\subsubsection{Density Forecasts: The Contribution of the Noncausal Specification}

The previous comparison held the underlying noncausal model fixed and assessed the contribution of the density estimator. We now examine the contribution of the model specification by comparing the MAR(1,1) with three leading univariate causal benchmarks from \citet{BaumeisterForecastingNatural2026}: an AR(1), an autoregression whose lag order is selected by the AIC, and exponential smoothing. Like the MAR(1,1), these models are linear and rely exclusively on the price series. The comparison therefore focuses on the specification rather than on differences in the information set or estimator flexibility.

The three causal benchmarks are implemented exactly as in the energy forecasting literature, that is, with Gaussian innovations,
$\varepsilon_t \overset{\text{i.i.d.}}{\sim} \mathcal{N}(0, \sigma^2)$. Let $(c_j)_{j\ge 0}$ denote the moving-average coefficients of the model. The $h$-step-ahead forecast error is $\sum_{j=0}^{h-1} c_j \varepsilon_{t+h-j}$, a linear combination of Gaussian innovations, and therefore itself Gaussian. The predictive density is available in closed form,\footnote{The conditioning vector $\mathbf{X}_t$ collects the lags each model requires: the last observation for the AR(1), the at most six lags selected by the AIC for the AR(AIC), and, for exponential smoothing, the entire history, summarised recursively by the exponentially weighted level.}
\begin{equation}
X_{t+h} \mid \mathbf{X}_t \; \sim \; \mathcal{N}\!\left(\hat{X}_{t+h \mid t}, \, \sigma_h^{2}\right),
\qquad
\sigma_h = \sigma \left( \sum_{j=0}^{h-1} c_j^{2} \right)^{1/2},
\label{eq:exact_gaussian_density}
\end{equation}
centred at the model's point forecast $\hat{X}_{t+h \mid t}$. The autoregressive recursion gives $c_j$ for the two AR specifications, and $c_0 = 1$, $c_j = 1-\theta$ for exponential smoothing.\footnote{The smoothing constant is fixed at $\theta = 0.8$, as in \cite{BaumeisterForecastingNatural2026}.} At every forecast origin we re-estimate the autoregressive coefficients by least squares and set $\sigma$ to the standard deviation of the corresponding residuals, on the same real-time vintage as the MAR(1,1). All four specifications are re-estimated at every forecast origin using the same real-time vintage, so the comparison subjects them to a common information set and re-estimation protocol. What separates these benchmarks from the noncausal predictive density are precisely the two ingredients that characterize the latter: the direction of the dynamics and the heavy-tailed nature of the innovations. If neither contributes beyond what a causal Gaussian representation already captures, the resulting closed-form densities should perform similarly to those of the MDN.

\ExecuteMetaData[outputs/applications.tex]{gas-comparison-training-dgp}

Table~\ref{tab:gas_comparison_training_dgp} indicates that the noncausal specification materially improves forecast accuracy. The MDN ranks first in 55 of the 63 metric-by-horizon cells, compared with 6 for the AR(1), 2 for the AR(AIC), and none for exponential smoothing, and belongs to the $MCS_{75\%}$ in 53 cells. Its advantage is most pronounced for the log-probability and censored likelihood scores, where it leads at all nine horizons, and for the quantile score, where it leads in 16 of the 18 cells. The causal benchmarks remain more competitive under the CDE loss and CRPS, for which the MDN leads in 6 and 7 of the 9 cells, respectively. The ranking also varies with the forecast horizon. At $h=1$, the Gaussian AR(1) leads in 5 of the 7 cells. From $h=3$ onward the ordering reverses: the MDN leads in six or seven cells at every horizon and belongs to the MCS in as many, except at $h=24$, where it is retained in three. Because all models use the same price series and re-estimation protocol, this advantage is attributable to the specification itself. The noncausal, heavy-tailed model yields predictive distributions that improve on those of the causal Gaussian benchmarks.

\afterpage{\FloatBarrier}
\subsubsection{Point Forecasts: Comparison with the Energy Forecasting Benchmarks}

A natural question is whether the improvements in density forecasting extend to point forecasting, although the predictive density remains our primary object of interest. Table~\ref{tab:mspe_comparison} compares point forecasts obtained from the median of the MDN predictive density with the models evaluated by \citet{BaumeisterForecastingNatural2026}. Forecast accuracy is measured by MSPE ratios relative to the no-change forecast, $\hat X_{t+h}=X_t$. The comparison includes the univariate models considered above and the six-variable Bayesian VARs of \citet{BaumeisterForecastingNatural2026}, which add dry natural gas production, natural gas rig counts, working gas inventories, capacity utilization, and the Residential Energy Demand Temperature Index to the real price.

We also evaluate point-forecast accuracy separately at center and tail origins. In what follows, tail origins are defined as in the previous subsections: those at which the last conditioning value exceeds the real-time 90th percentile of the prices available in that origin's vintage.

The regional results reveal a pronounced state dependence. At tail origins, the MDN is the most accurate of all models, including the multivariate benchmarks, at eight of the nine horizons. Its MSPE ratio reaches 0.692 at $h=12$; exponential smoothing leads only at $h=18$. At center origins, by contrast, the MDN does not lead: the AR(1) performs best among the univariate models through $h=6$, and the AR(AIC) leads thereafter.

The $MCS_{75\%}$, computed over the univariate models only, retains the MDN at every horizon both overall and at tail origins, and at all but $h=18$ in the center. It never reduces to a single model. The low power of the test is expected: the comparison rests on 325 origins overall and 73 at tail origins, and squared errors of a heavy-tailed price yield imprecisely estimated loss differentials.

The comparison with the multivariate models is more nuanced. Despite relying only on the price series, the MDN attains a lower overall MSPE ratio than the BVAR(AIC) at seven of the nine horizons and than the BVAR(1) at $h=21$. At tail origins, it outperforms both BVAR specifications at every horizon. The ordering reverses in the center, where the additional predictors used by the BVARs appear more informative. Their overall advantage is therefore concentrated in relatively normal market conditions, whereas the MDN performs particularly well when the current price is in the upper tail.

\ExecuteMetaData[outputs/applications.tex]{mspe-comparison}

These findings are consistent with the state dependence of the noncausal predictive distribution. The point-forecast gains are concentrated at tail origins, where the MAR(1,1) captures dynamics that are more difficult to represent using the causal alternatives considered by \citet{BaumeisterForecastingNatural2026}.

Overall, despite not being designed specifically for point forecasting, the MDN ranks among the leading univariate methods and remains competitive with multivariate models that use substantially richer information sets.

\section{Conclusion}
\label{s6}

This paper develops a practical framework for density forecasting with mixed causal--noncausal ARMA processes. The approach separates model estimation from density approximation: a noncausal process is first estimated, and a tail-aware Mixture Density Network is then trained on trajectories generated from the fitted model. The network combines skewed-$t$ mixture components, adaptive weighting of extreme conditioning states, and post-hoc recalibration.

The Monte Carlo analysis evaluates the estimated densities against three complementary benchmarks: the closed-form predictive density available for the Cauchy MAR(0,1), realized outcomes assessed with proper scoring rules, and theoretical conditional moments when they exist. Across the specifications and forecast horizons considered, the MDN generally provides the most accurate approximation. Its advantage is most pronounced in the tails, where noncausal predictive distributions display their strongest state dependence and may exhibit distinct continuation and collapse components.

The real-time application to U.S. natural gas prices confirms the practical relevance of these gains. Holding the fitted MAR(1,1) specification fixed, the MDN outperforms the alternative density estimators overall and performs particularly well under scoring rules that emphasize the upper tail. Comparisons with causal Gaussian benchmarks further indicate that the noncausal, heavy-tailed specification contributes materially to forecast accuracy. Although designed for density forecasting, the approach also delivers competitive point forecasts, with its largest relative gains arising when the current natural gas price is already in the upper tail of the price distribution.

More broadly, the results illustrate how machine-learning methods can complement structurally motivated time-series models. Rather than learning directly from the observed series or replacing the underlying economic model, the neural network approximates the predictive distribution implied by the fitted process. This division of labor preserves the interpretation of noncausal dynamics while making their forecasts operational. Future research could extend the framework to multivariate processes and incorporate parameter and specification uncertainty more directly into the learned predictive density.

\clearpage
\bibliographystyle{oup-abbrvnat}
\bibliography{biblio}

\end{document}